\documentclass[twocolumn]{aastex62}

\usepackage{bm}
\usepackage{url}
\usepackage{amsmath}
\usepackage{color}
\usepackage{graphicx}
\usepackage{slashbox}

\received{\today}
\revised{\today}
\accepted{\today}
\submitjournal{ApJ}

\shorttitle{Scattering-induced intensity reduction}
\shortauthors{Ueda et al.}

\begin{document}

\title{
Scattering-induced intensity reduction: large mass content with small grains in the inner region of the TW Hya disk
}

\correspondingauthor{Takahiro Ueda}
\email{takahiro.ueda@nao.ac.jp}

\author[0000-0003-4902-222X]{Takahiro Ueda}
\affil{National Astronomical Observatory of Japan, Osawa 2-21-1, Mitaka, Tokyo 181-8588, Japan}

\author[0000-0003-4562-4119]{Akimasa Kataoka}
\affil{National Astronomical Observatory of Japan, Osawa 2-21-1, Mitaka, Tokyo 181-8588, Japan}

\author[0000-0002-6034-2892]{Takashi Tsukagoshi}
\affil{National Astronomical Observatory of Japan, Osawa 2-21-1, Mitaka, Tokyo 181-8588, Japan}



\begin{abstract}
Dust continuum observation is one of the best methods to constrain the properties of protoplanetary disks.
Recent theoretical studies have suggested that the dust scattering at the millimeter wavelength potentially reduces the observed intensity, which results in an underestimate in the dust mass.
We investigate whether the dust scattering indeed reduces the observed continuum intensity by comparing the ALMA archival data of the TW Hya disk at Band 3, 4, 6, 7 and 9 to models obtained by radiative transfer simulations.
We find that the model with scattering by 300 ${\rm \mu m}$-sized grains well reproduces the observed SED of the central part of the TW Hya disk while the model without scattering is also consistent within the errors of the absolute fluxes.
To explain the intensity at Band 3, the dust surface density needs to be $\sim$ 10 ${\rm g\,cm^{-2}}$ at 10 au in the model with scattering, which is 26 times more massive than previously predicted. 
The model without scattering needs 2.3 times higher dust mass than the model with scattering because it needs lower temperature.
At Band 7, scattering reduces the intensity by $\sim$ 35\% which makes the disk looks optically thin even though it is optically thick.
Our study suggests the TW Hya disk is still capable of forming cores of giant planets at where the current solar system planets exist.
\end{abstract}


\section{Introduction}
(Sub-)Millimeter continuum observations have allowed us to examine the properties of protoplanetary disks.

From the continuum emission, the dust mass can be estimated for a given dust opacity and temperature with assuming that the disk is optically thin \citep{Beckwith1990,Andrews2005,Andrews2013,Ansdell2016,Cieza2019}.
Furthermore, if we look at a disk with multi-wavelength, the spectral slope of the intensity allows us to estimate the dust size (e.g., \citealt{Calvet2002,Draine2006}).
However, if the disk is optically thick, it is hard to obtain the information about the dust disk below the photosphere.
Even though the inner region where the current solar system planets exist ($\lesssim 30$ au) is of most importance in understanding how solar-system analogs form, inner region of disks would be optically thick at (sub-)millimeter wavelengths since protoplanetary disks are supposed to have higher surface density at more inner radius (e.g., \citealt{Lynden1974,Hayashi1981}).

However, the recent high angular resolution observations from the Disk Substructures at High Angular Resolution Project (DSHARP, \citealt{Andrews2018}) revealed that most of the DSHARP disks have optical depths less than unity even in the inner region \citep{Huang2018}.
One explanation for these low observed optical depth is the self-scattering of the thermal emission from dust grains (\citealt{Birnstiel2018,Liu2019,Zhu2019,Carrasco2019}; see also Appendix C in \citealt{Miyake1993}).
\cite{Zhu2019} showed that if the disk is optically thick, the dust scattering may reduce the observed optical depth by orders of magnitude and the disk looks optically thin even if the true optical depth is sufficiently larger than unity.
This is because scattering makes the effective free path of emitted photons shorter than that without scattering so that we can only see the upper layer of the disk.

\begin{figure*}[ht]
\begin{center}
\includegraphics[scale=0.41]{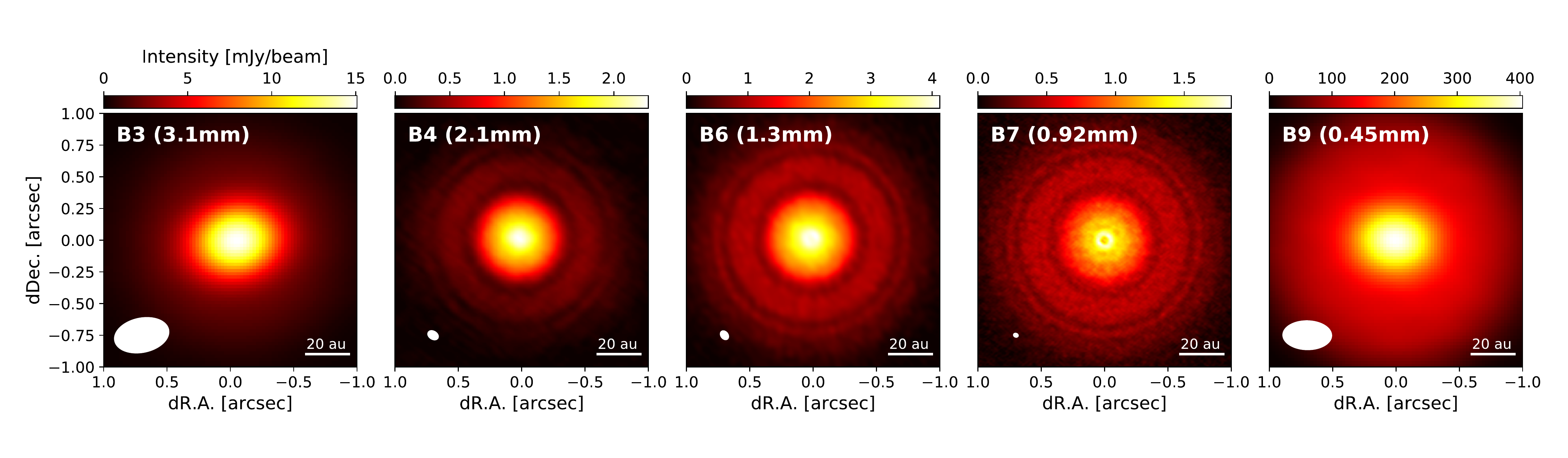}
\caption{
ALMA continuum images of the TW Hya disk at Band 3, 4, 6, 7 and 9 from left to right.
The synthesized beam sizes at Band 3, 4, 6, 7 and 9 are $0.43\times0.27$, $0.088\times0.062$, $0.075\times0.055$, $0.036\times0.029$ and $0.38\times0.23$ arcsec, respectively.
}
\label{fig:images}
\end{center}
\end{figure*}

The dust scattering also affects the spectral index $\alpha$ by reducing the observed intensity \citep{Liu2019,Zhu2019}.
The spectral index $\alpha$ is a power law index of the intensity $I$ as a function of the frequency $\nu$, which has a value higher than 2 for the optically thin disk and as low as 2 for the optically thick disk in the Rayleigh-Jeans limit.
However, some disks have shown spectral indices anomalously lower than 2 (e.g.,\citealt{Tsukagoshi2016,Nomura2016,Liu2017,Huang2018a,Dent2019}).
\cite{Liu2019} demonstrated that the optically thick disk can have a spectral index lower than 2 at the millimeter wavelength because of the intensity reduction by scattering and showed that it would account for the low spectral index in the inner region of the TW Hya disk reported by \cite{Tsukagoshi2016}.

Recent ALMA polarimetric observations have shown that the dust scattering indeed occurs in some disks \citep{Kataoka2016a,Kataoka2016b,Bacciotti2018,Hull2018,Ohashi2018,Dent2019}.
The polarization caused by the dust scattering shows the polarization pattern where the polarization vector is parallel to the minor axis of the disk.
From the polarization degree at the observing wavelength, we can put a constrain on the dust size since the scattering behavior is sensitive to the ratio of the dust size to the observing wavelength $\lambda$ \citep{Kataoka2015}.
These observations indicate that the dust scattering might affect not only the polarization morphology but also the intensity of the continuum emission.

In this paper, we analyze the ALMA archive data of the continuum observations of the TW Hya disk at Band 3, 4, 6, 7 and 9 to investigate whether the dust scattering indeed affects the observed intensity in the inner region of the disk.
In Section \ref{sec:2}, we analyze the ALMA archive data and show that the intensities at Band 7 and 9 are lower than the intensity extrapolated from longer wavelengths with the spectral slope of 2.
In Section \ref{sec:3}, we perform radiative transfer simulations and demonstrate that the observed SED is well explained by the model with 300 ${\rm \mu m}$ sized grains with including scattering.
The properties of the inner region of the TW Hya disk are discussed in Section \ref{sec:diss} and the summary is in Section \ref{sec:summary}.

\section{Millimeter SED of the central part of the TW Hya disk} \label{sec:2}
We investigate the multi-wavelength observations of the inner part of the TW Hya disk where the previous studies have suggested that it is optically thick at Band 4 and 6 (e.g., \citealt{Tsukagoshi2016}).
Because the lowest spatial resolution in our data set ($\sim 20$ au, corresponding to $\sim0.34$ arcsec) is not enough to discuss the detailed radial structures, we investigate the Spectral Energy Distribution (SED) of the central part of the disk as shown in the following sections.

\subsection{ALMA data and analysis}
In this study, we analyze the public images of the continuum emission of the TW Hya disk at Band 3, 4, 6, 7, and 9 taken with ALMA.
The Band 4 and 6 images are adopted from \citet{Tsukagoshi2016}, and the Band 7 image is from \citet{Tsukagoshi2019}.
For the detailed calibration process of Band 4, 6 and 7 images, we refer readers to \citet{Tsukagoshi2016} and \citet{Tsukagoshi2019}.
For the Band 3 and 9 images, we downloaded the product images provided by ALMA from the ALMA data archive system (2016.1.00229.S and 2012.1.00422.S, PI: Edwin Bergin).
The observed visibilities were reduced and calibrated using the Common Astronomical Software Application (CASA) package \citep{McMullin2007}. 
The initial flagging of the visibilities and the calibrations for the bandpass characteristics, complex gain, and flux scaling were performed using the pipeline scripts provided by ALMA.
After flagging the bad data, the corrected data were concatenated and imaged by CLEAN. 
The CLEAN map was created by adopting Briggs weighting with a robust parameter of 0.8 for Band 3 and 0.5 for Band 9.
After that, self-calibration was applied for the data set.

Figure \ref{fig:images} shows the continuum images of the TW Hya disk at each band.
The angular resolution of each observation is $0.43\times0.27$, $0.088\times0.062$, $0.075\times0.055$, $0.036\times0.029$ and $0.38\times0.23$ arcsec for Band 3, 4, 6, 7 and 9, respectively.
Owing to the high spatial resolution, the images at Band 4, 6 and 7 clearly show the concentric ring and gap-like structures which have been already reported (e.g., \citealt{Andrews2016,Tsukagoshi2016}).
In contrast, the Band 3 and 9 images have low angular resolution so that the substructures are not seen.
The maximum recoverable scale of the obtained data is 5.0, 18, 12, 8.8 and 5.1 arcsec for Band 3, 4, 6, 7 and 9, respectively.
Since the diameter of the dust disk around TW Hya is $\sim$ 140 au, corresponding to the angular scale of $\sim$ 2.3 arcsec, our data sufficiently recovers the whole disk emission.

\subsection{SED analysis}
We focus only  on the inner region within the radial distance of $\sim$ 10 au where the previous study showed that the spectral index between Band 4 and 6 is lower than 2 \citep{Tsukagoshi2016}.
In order to investigate each data with the same beam size, the observed images at Band 4, 6, 7 and 9 are re-convolved with the beam size of Band 3 observation ($0.43\times0.27$ arcsec) which is equivalent to the spatial resolution of $26\times16$ au for the distance of the target from the Earth ($\approx$ 60 pc, \citealt{Gaia2016}).

\begin{figure}[ht]
\begin{center}
\includegraphics[scale=0.48]{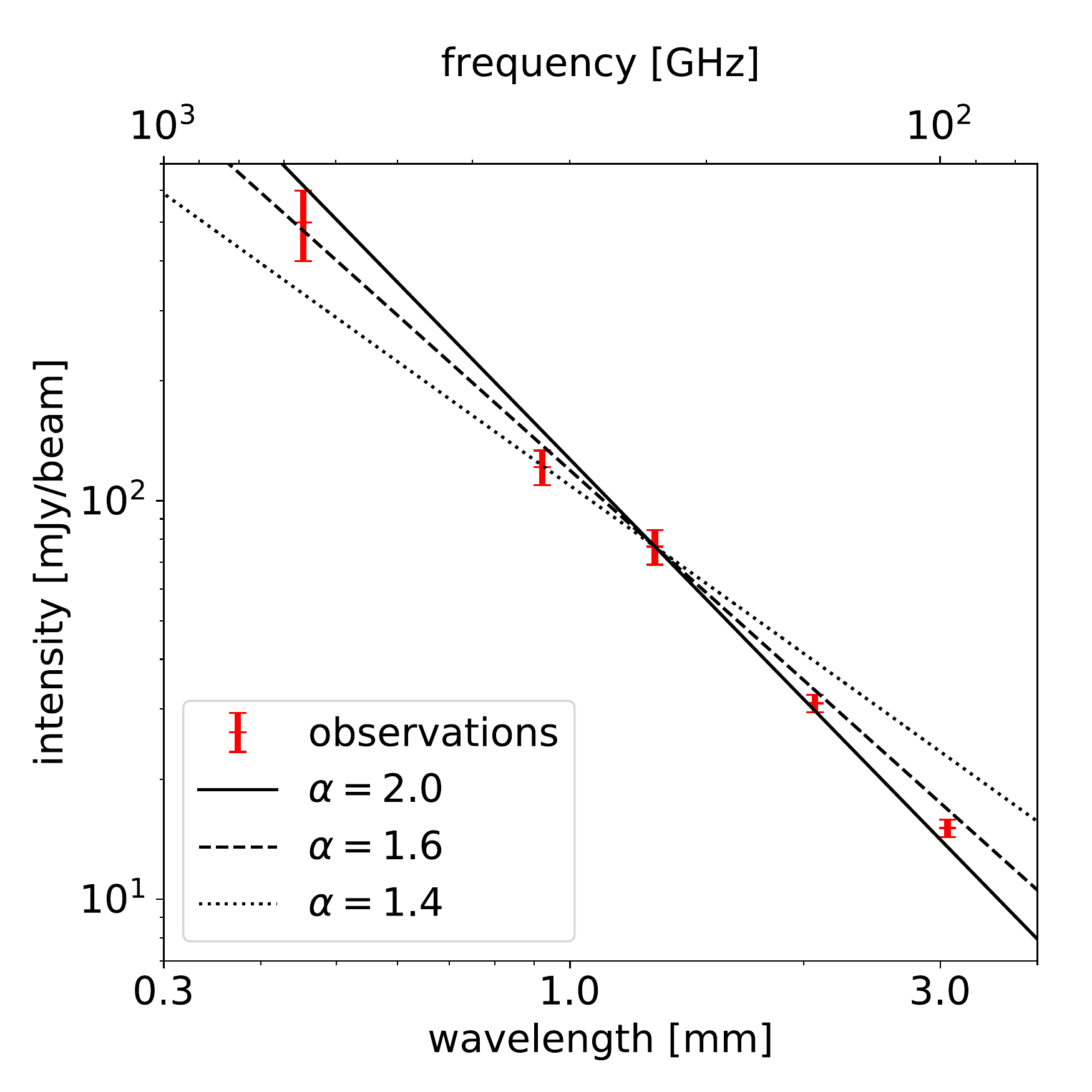}
\caption{
Intensities at the center of the re-convolved images at different observing wavelengths.
For reference, the intensity profiles following the spectral index of 2, 1.6 and 1.4 are denoted by the solid, dashed and dotted line, respectively.
The uncertainty in the absolute intensity is set to be 5\% for Band 3 and 4, 10\% for Band 6 and 7 and 20\% for Band 9.
}
\label{fig:sed-inside}
\end{center}
\end{figure}

\begin{table}[ht]
  \begin{center}
  \begin{tabular}{|c|c|c|c|}
  \hline
Wavelength  & Intensity & Noise & Uncertainty\\
 $[{\rm \mu m}]$ & [mJy/beam] & [mJy/beam] & [mJy/beam]\\
\hline\hline
453.5 (B9) & 499 & 1.1 &$\pm99.8$\\\hline
921.3 (B7) & 122 & 0.24 &$\pm12.2$\\\hline
1287 (B6) & 76.7 & 0.12 &$\pm7.67$\\\hline
2068 (B4) & 31.0 & 0.091 &$\pm1.55$\\\hline
3064 (B3) & 15.1 & 0.025 &$\pm0.755$\\ \hline
  \end{tabular}
  \caption{Observed intensity at the center of the images and the 1$\sigma$ noise level at different observing wavelengths.
  The uncertainty in the absolute intensity is set to be 5\% for Band 3 and 4, 10\% for Band 6 and 7 and 20\% for Band 9.
  }
  \end{center}
  \label{table:1}
\end{table}

The intensities at the center of the re-convolved images at each band are shown in Figure \ref{fig:sed-inside} and summarized in
Table \ref{table:1}.
Since the spatial resolution of $26\times16$ au corresponds to the effective beam radius of 10 au, the intensity at the center of the images traces mainly the inner disk within the radius of $\sim$ 10 au.
In Figure \ref{fig:sed-inside}, we plot error bars corresponding to 5\% of the absolute intensity for Band 3 and 4, 10\% for Band 6 and 7 and 20\% for Band 9. 
The errors are potentially caused in the flux calibration process and quoted from ALMA official observing guide.
In Appendix \ref{sec:ap1}, we examined the time variation of the flux density of the amplitude calibrators to see the actual uncertainty.
We confirmed that the actual uncertainty might be potentially larger than the official value but has little impact on our conclusion.

Figure \ref{fig:sed-inside} shows that the observed SED does not follow a single power-law profile.
The spectral slope through Band 3 to 6 can be explained by the spectral index of 2 within the error.
The spectral index lower than 2 between Band 4 and 6 reported by \citet{Tsukagoshi2016} is also consistent with our analysis.
In contrast, the intensities at Band 7 and 9 are lower than the intensity extrapolated from longer wavelengths with the spectral slope of 2. 
The spectral index between Band 6 and 7 is $\approx 1.4$ which is significantly lower than 2.
The anomalously low spectral index around Band 7 is consistent with the previous studies within the error (\citealt{Nomura2016,Huang2018a}).
In addition, the spectral index between Band 6 and 9 is $\approx1.6$ which is also lower than 2 but higher than that between Band 6 and 7.

The observed low spectral indices through Band 6 to 9 indicate that the scattering-induced intensity reduction occurs and it is the most effective at Band 7 in the inner region of the disk.
To explain the observed SED of the inner part of the TW Hya disk, we perform radiative transfer simulations in the following section.

\section{Radiative transfer modeling} \label{sec:3}

\subsection{Modeling concepts}
In this subsection, we briefly describe the concept of how we model the simulated disks.
\begin{figure}[ht]
\begin{center}
\includegraphics[scale=0.48]{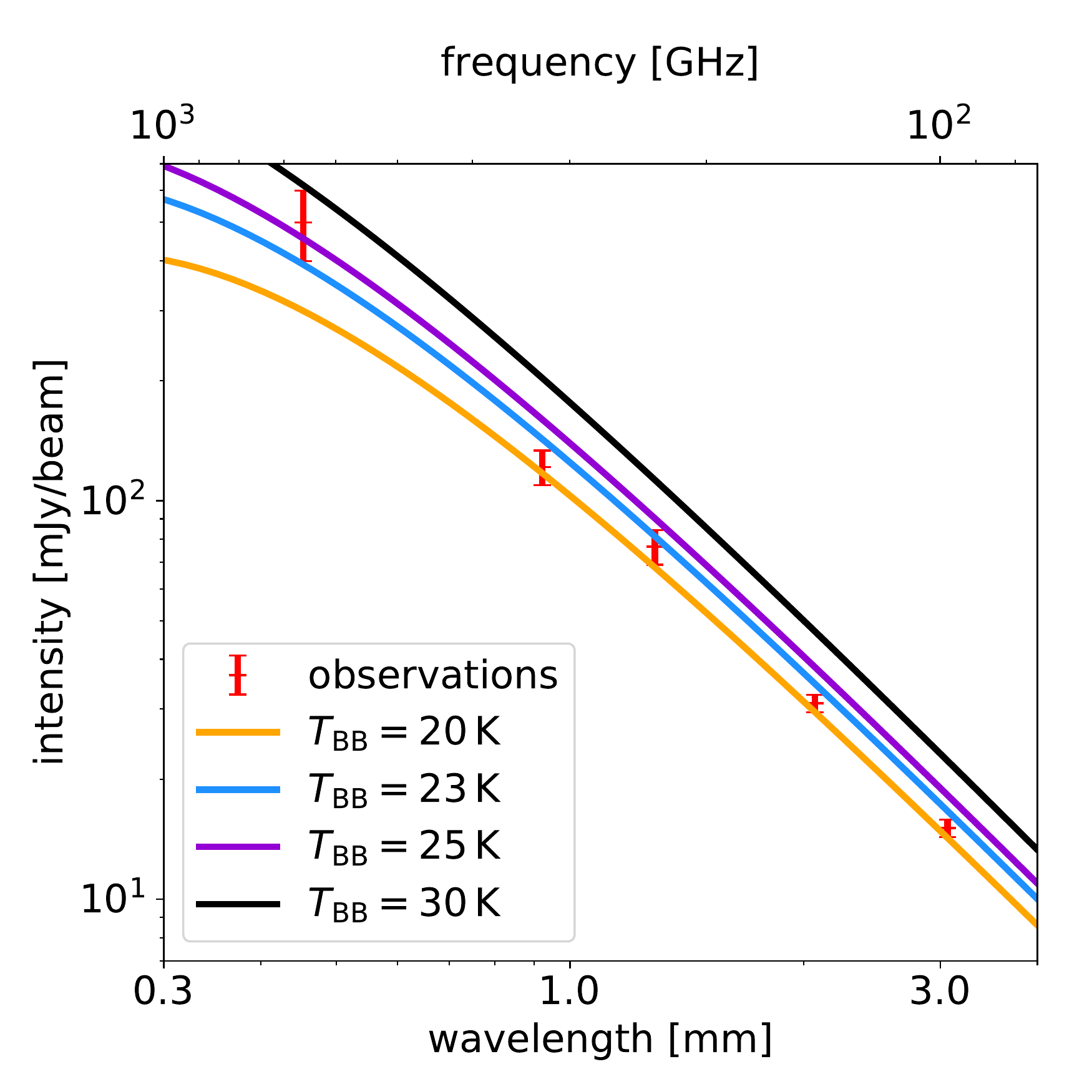}
\caption{
Blackbody curves at dust temperatures of 20 K (orange), 23 K (light blue), 25 K (purple) and 30 K (black).
The red points show the intensities at the center of the observed images.
}
\label{fig:sed-inside-bb}
\end{center}
\end{figure}
As a first step, we compare the black body curves with different temperatures $T_{\rm BB}$ to the observed SED in Figure \ref{fig:sed-inside-bb}.
Figure \ref{fig:sed-inside-bb} shows that the observed SED is similar to the black body curve with temperature of $\sim$ 23 K.
Because the disk is likely optically thick at Band 9 and the observed intensity at Band 9 can be reproduced with $T_{\rm BB}\gtrsim23 $K, the lower limit of the disk temperature can be estimated as $\sim$ 23 K.
If the scattering-induced intensity reduction occurs at Band 9, the disk temperature could be higher than 23 K.

The observed intensity at Band 3 is almost consistent with the black body curve with $T_{\rm BB}=23 $K or slightly lower.
Therefore, if scattering is not effective at Band 3 and the temperature is 23 K, the disk is (marginally) optically thick even at Band 3. If the temperature is higher than 23 K, the disk should be optically thin or scattering should be effective.

In this work, we focus only on the emission from the center of the disk images which is observed with the beam radius of $\sim$ 10 au.
Since the emission would be dominated by the outer part within the observing beam, these estimated temperature and optical thickness would trace values at the outer edge of the beam ($\sim$ 10 au from the center).
From these analysis, we model the disk so that the disk is marginally optically thin/thick at Band 3 and temperature is higher than 20 K at 10 au.
Detailed setup will be shown in the following section.

\subsection{Radiative transfer setup}
\begin{figure}[ht]
\begin{center}
\includegraphics[scale=0.41]{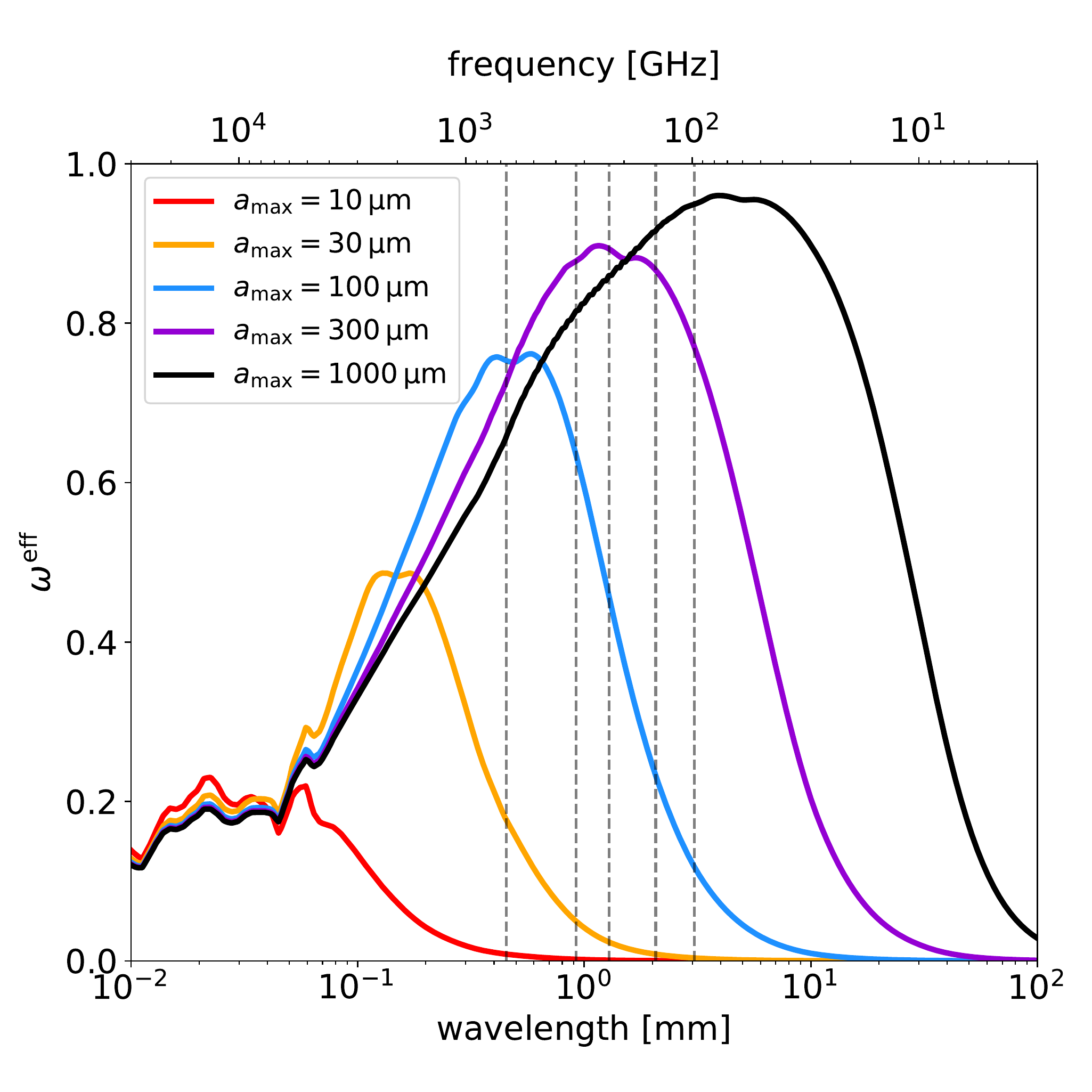}
\caption{
Effective scattering albedo $\omega_{\rm eff}$ as a function of the wavelength for the different maximum dust sizes.
The gray vertical dashed lines show the observing wavelengths at Band 3, 4, 6, 7 and 9 from right to left.
}
\label{fig:albedo}
\end{center}
\end{figure}

In order to investigate the properties of the inner part of the TW Hya disk, radiative transfer simulations are performed with the Monte Carlo radiative transfer code RADMC-3D \citep{RADMC}.
For dust opacities, we use the DSHARP dust optical constants published in \citet{Birnstiel2018} (see also \citealt{Henning1996,Draine2003,Warren2008} for the optical constants of each dust component).
The dust grains are assumed to be spherical compact grains with the material density of 1.675 $
{\rm g~cm^{-3}}$.
The dust size distribution is assumed to be a power-law distribution ranging from 0.1 ${\rm \mu m}$ to $a_{\rm max}$ with the power-law index of 3.5.
To treat full anisotropic scattering, the M\"{u}eller matrices are calculated using the Mie theory, specifically Bohren-Huffman program \citep{Bohren1983}.
Figure \ref{fig:albedo} shows the effective scattering albedo $\omega_{\rm eff}$ for the different maximum dust sizes. 
The effective scattering albedo is defined as
\begin{eqnarray}
\omega_{\rm eff}=\frac{\kappa_{\rm sca}^{\rm eff}}{\kappa_{\rm abs}+\kappa_{\rm sca}^{\rm eff}},
\label{eq:albedo}
\end{eqnarray}
where $\kappa_{\rm abs}$ is the absorption opacity and $\kappa_{\rm sca}^{\rm eff}$ is the scattering opacity considering the effect of forward scattering.
The values of the effective scattering albedo at each observing band are summarized in Table \ref{table:2}.
At (sub-)millimeter wavelength, the effective scattering albedo increases with the wavelength when the wavelength is shorter than $\sim 2\pi a_{\rm max}$ and has the maximum at around $\lambda \sim 2\pi a_{\rm max}$ and drops off when $\lambda > 2\pi a_{\rm max}$.

\begin{table}[ht]
  \begin{center}
  \begin{tabular}{|c||c|c|c|c|c|}
  \hline
\backslashbox{$a_{\rm max}$}{Band} & 3  & 4 & 6  & 7 & 9\\
\hline\hline
$10~{\rm \mu m}$ & 0.00015 & 0.00033 & 0.00094 & 0.0020 & 0.0085 \\\hline
$30~{\rm \mu m}$ & 0.0037 & 0.0084 & 0.024 & 0.050 & 0.18 \\\hline
$100~{\rm \mu m}$ & 0.12 & 0.23 & 0.46 & 0.63 & 0.75 \\\hline
$300~{\rm \mu m}$ & 0.77 & 0.87 & 0.89 & 0.88 & 0.73 \\\hline
$1000~{\rm \mu m}$ & 0.95 & 0.92 & 0.86 & 0.82 & 0.66 \\ \hline
  \end{tabular}
  \caption{Effective scattering albedo $\omega_{\rm eff}$ at each observing band.}
  \end{center}
  \label{table:2}
\end{table}

In our simulations, the dust surface density is assumed to follow a simple power law profile:
\begin{eqnarray}
\Sigma_{\rm d}= \Sigma_{\rm 10}\left( \frac{r}{\rm 10~au} \right)^{-0.5},
\label{eq:surfacedensity}
\end{eqnarray}
where $r$ is the mid-plane distance from the central star and $\Sigma_{\rm 10}$ is the dust surface density at 10 au.
The inner edge of the disk is set to 1 au to mimic the presence of a inner cavity in the TW Hya disk.
We truncate the disk at 50 au which is enough far from the center of the disk to not affect the intensity at the center when the intensity is convolved with the observing beam size ($\sim$ 20 au).
The dust surface density is related to the vertical absorption optical depth at Band 3 at 10 au as,
\begin{eqnarray}
\tau_{10}=\kappa_{\rm B3}\Sigma_{\rm 10},
\label{eq:tau}
\end{eqnarray}
where $\kappa_{\rm B3}$ is the absorption opacity at the observing frequency of Band 3.
Using the dust surface density, the dust volume density $\rho_{\rm d}$ is calculated as
\begin{eqnarray}
\rho_{\rm d}=\frac{\Sigma_{\rm d}}{\sqrt{2\pi h_{\rm d}}}\exp{ \left(-\frac{z^{2}}{2h_{\rm d}^{2}}\right) },
\label{eq:volumedensity}
\end{eqnarray}
where $z$ is the vertical height from the mid plane and $h_{\rm d}$ is the scale height of the dust disk,
\begin{eqnarray}
h_{\rm d}= 0.63\left( \frac{r}{\rm 10\, au} \right)^{1.1}~{\rm au}.
\label{eq:scaleheight}
\end{eqnarray}
The temperature profile is assumed to be
\begin{eqnarray}
T(r,\phi,z)= T_{\rm 10}\left( \frac{r}{\rm 10~au} \right)^{-0.4},
\label{eq:temperature}
\end{eqnarray}
where $\phi$ is the azimuthal angle and $T_{\rm 10}$ is the temperature at 10 au.
The disk inclination is set to be $7^{\circ}$ based on \citet{Qi2004}.
The simulated images are convolved with the observing beam size.

In the following subsections, we show the comparison between the ALMA observations and results of the radiative transfer simulations with and without scattering.

\subsection{Step 1 - no scattering case}
\label{sec:nosca}
As shown in Figure \ref{fig:sed-inside}, the intensities at short wavelengths are lower than the intensity extrapolated from longer wavelengths with the spectral slope of 2.
At short wavelengths, the observing wavelength would be close to the peak wavelength of the black body emission from the disk so that the black body emission might be deviated from and lower than the Rayleigh-Jeans limit (e.g., Figure 5 of \citealt{Kim2019}).

\begin{figure*}[ht]
\begin{center}
\includegraphics[scale=0.42]{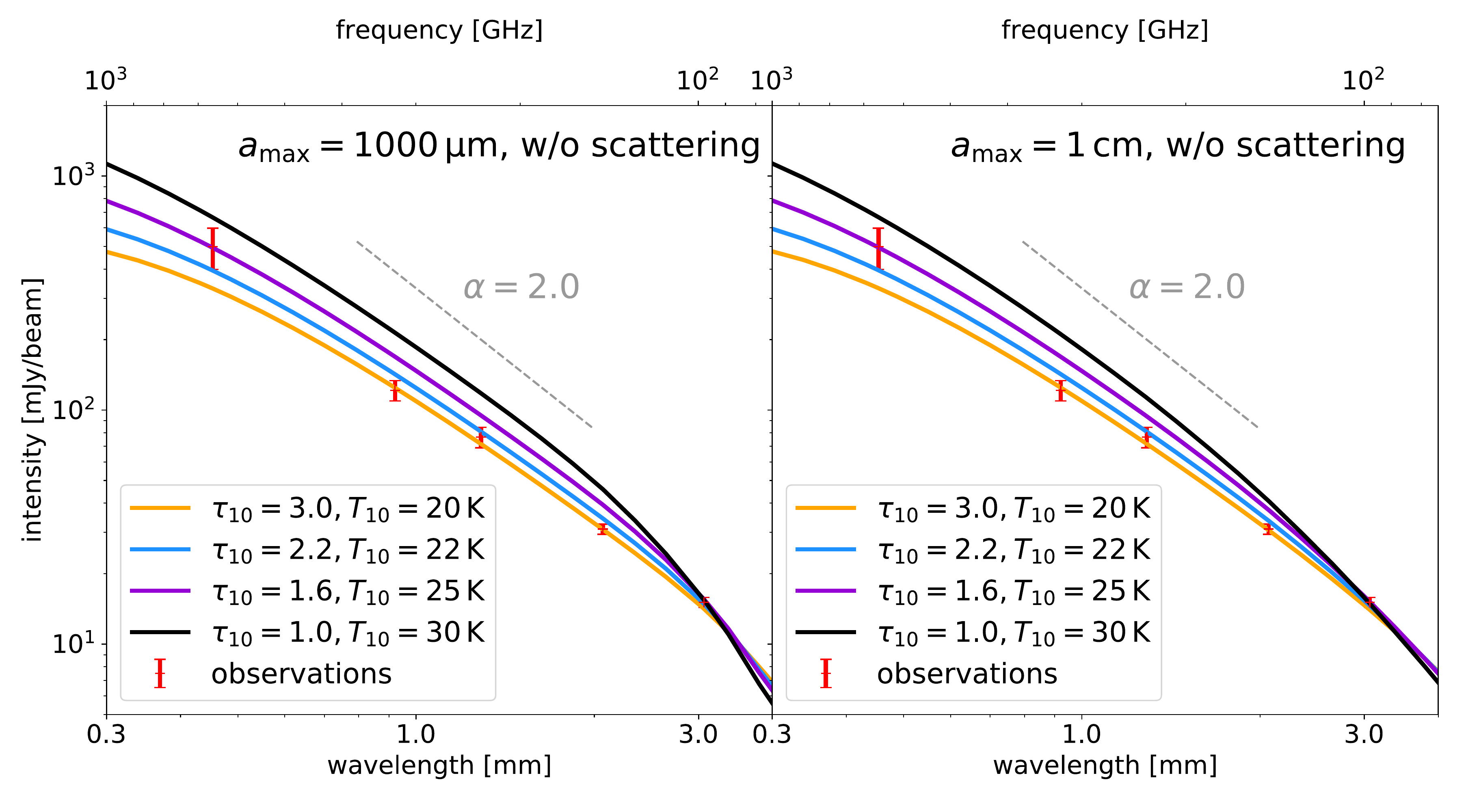}
\caption{
Intensity at the center of the disk images simulated without scattering.
The orange, light blue, purple and black solid lines show the simulated intensity for the different disk models .
The maximum dust size is set to be $1000~{\rm \mu m}$ and $1~{\rm cm}$ in left and right panels, respectively.
The red points show observed values.
The gray dashed line denotes the spectral index of 2.
}
\label{fig:sed-temp}
\end{center}
\end{figure*}

In order to investigate whether the observed low intensity at high frequency can be explained simply by the deviation from the Rayleigh-Jeans limit, we first show the results of radiative transfer simulations without scattering in Figure \ref{fig:sed-temp}.
In Figure \ref{fig:sed-temp}, we change the disk temperature, and the disk mass (i.e., $\tau_{10}$) is set for each temperature model so that the observed intensity at Band 3 is reproduced.
If $T_{10}=20$ K, the simulated intensity profile is similar to the observed one between Band 3 and 7 but is significantly lower at Band 9.
If $T_{10}$ is higher than 30 K, the model overpredicts the intensity at Band 4--9.
Although the observed intensity profile is barely reproduced by the model with $T_{10}= 22$ K, the model intensities are the lower limit of the Band 9 observation and higher at Band 6.
If $T_{10}$ is higher than 22 K, the intensity at observing wavelength shorter than Band 4, especially at Band 4, 6 and 7,  is significantly higher than the observed value.
Since the disk is optically thick at $\lambda\lesssim3~{\rm mm}$ in all of these models, the impact of dust size arises only at $\lambda\gtrsim3~{\rm mm}$.
Therefore we conclude that,  if scattering is not taken into account, it is difficult to explain the observed intensity profile except for the model with $\tau_{10}=2.2$ and $T_{10}=22$ K.

\subsection{Step 2 - turning on the scattering effect}
\label{sec:best}
\begin{figure}[ht]
\begin{center}
\includegraphics[scale=0.42]{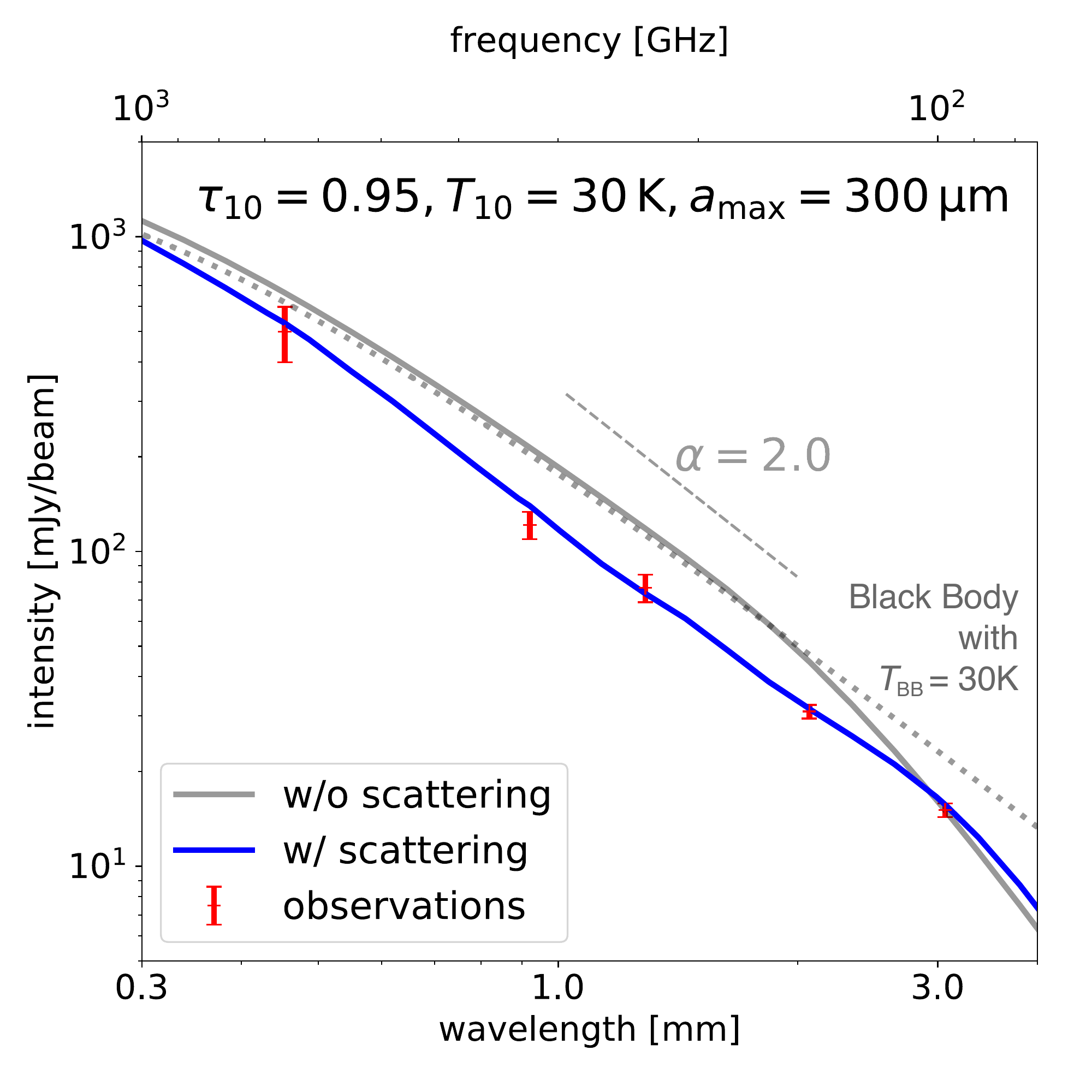}
\caption{
Intensity at the center of the disk images.
The blue and gray solid lines show the simulated intensity for the maximum dust size of $300~{\rm \mu m}$ with and without scattering, respectively.
The vertical absorption optical depth at 10 au is set to be 0.95 at Band 3 and the temperature at 10 au is set to be 30 K.
The red points show observed values.
The gray dotted line shows the black body curve at dust temperature of 30 K.
The gray dashed line denotes the spectral index of 2.
}
\label{fig:fiducial}
\end{center}
\end{figure}
As the observed SED suggests that the dust scattering is the most effective at Band 7, we first perform the radiative transfer simulation with turning on scattering and set the maximum dust radius to be 300 ${\rm \mu m}$ whose albedo has a peak around Band 6--7 (see Figure \ref{fig:albedo}).

Figure \ref{fig:fiducial} shows the comparison between the observed intensity and the simulated intensity of the model with the maximum dust radius of 300 ${\rm \mu m}$ with and without scattering.
In Figure \ref{fig:fiducial},  $\tau_{10}$ and $T_{10}$ are set to be 0.95 and 30 K, respectively.
We clearly see that the model with scattering well produces the observed intensities while the model without scattering overestimates.
As the scattering albedo of 300 ${\rm \mu m}$ grains has a peak around Band 6--7, scattering is the most effective at Band 6--7 and reduces the intensity at those bands by $\sim$ 35\%.

On the other hand, at the wavelength longer than 3 mm, the model with scattering shows the slightly higher intensity than the model without scattering.
This would be because, if the observed region is optically thin for the vertical direction but optically thick for the radial direction, scattered photons selectively escape to the vertical direction, which enhances the disk brightness compared to no-scattering case.
In this model, the dust scattering enhances the observed intensity by $\sim$ 20\% at the maximum compared to the model without scattering.
This intensity enhancement by scattering can be also seen in Figure 9 of \citet{Birnstiel2018} although they have not mentioned it.

\subsection{Step 3 - fine-tuning of grain size}
\begin{figure}[ht]
\begin{center}
\includegraphics[scale=0.42]{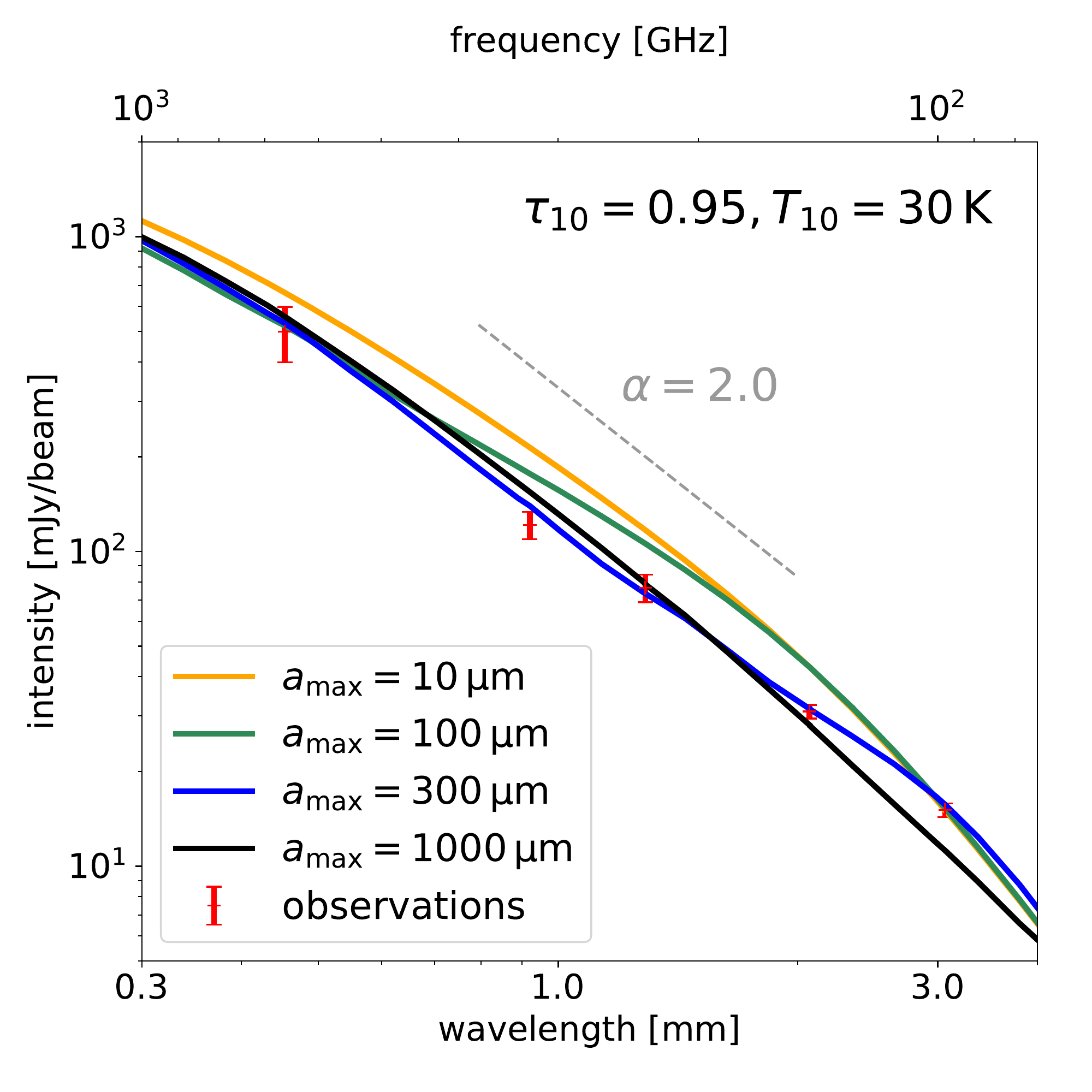}
\caption{
Intensity at the center of the disk images.
The orange, green, blue and black solid lines show the simulated intensity for the maximum dust size of $a_{\rm max}=10$, 100, 300 and 1000 ${\rm \mu m}$, respectively.
The vertical optical depth at 10 au is set to be 0.95 at Band 3 and the temperature at 10 au is set to be 30 K.
The red points show observed values.
The gray dashed line denotes the spectral index of 2.
}
\label{fig:dustsize}
\end{center}
\end{figure}
From the observed SED, one would expect that the dust scattering reduces the observed intensity only in the observing wavelength shorter than 1 mm.
This implies that dust grains needs to be smaller than 1 mm otherwise the dust scattering is effective even at longer observing wavelength.
In order to investigate the dependence of the SED on the dust size, we perform radiative transfer simulations with different dust sizes.

Figure \ref{fig:dustsize} shows simulated intensities with different dust sizes for the same disk temperature and mass with the model shown in Section \ref{sec:best}.
If the maximum dust radius is 10 ${\rm \mu m}$, scattering is not effective so that the simulated intensity is almost the same as that of the no-scattering model shown in Figure \ref{fig:fiducial}.
For the 100 ${\rm \mu m}$ grains, the effect of scattering shows up and more effective in the shorter observing wavelength since the scattering albedo of 100 ${\rm \mu m}$ grains monotonically decreases with the wavelength within the observing wavelengths.
These results indicate that dust grains smaller than 100 ${\rm \mu m}$ does not account for the observed intensity profile which has a dip around Band 7.
In contrast, if the dust size is larger than 1000 ${\rm \mu m}$, scattering is the most effective at Band 3 since the scattering albedo monotonically increases with the wavelength, which leads to an underestimate the intensity at Band 3.
Therefore the maximum dust radius of $\sim$ 300 ${\rm \mu m}$ whose scattering albedo has a peak around Band 6--7 is the best to explain the observed intensity profile.

\subsection{Modeling summary}
In this subsection, we briefly summarize our findings of the disk modeling.
First, if scattering is not taken into account, it is difficult to reproduce the observed SED except for the model with $\tau_{10}=2.2$ and $T_{10}=22$ K.  
The models with $T_{10}>22$ K overpredicts the intensity at Band 4--7, while the model with $T_{10}<22$ K underestimates the intensity at Band 9.
If the dust scattering is taken into account, the model with the maximum dust radius of 300 ${\rm \mu m}$ well reproduces the observed SED.
Since the dust scattering reduces the intensity by $\sim$ 35\% at Band 7, the disk needs higher temperature ($T_{10}=30$ K) than the model without scattering ($T_{10}=22$ K).
The optical depth is determined as $\tau_{10}=0.95$ since the disk is marginally optically thin at Band 3.

\begin{figure*}[ht]
\begin{center}
\includegraphics[scale=0.60]{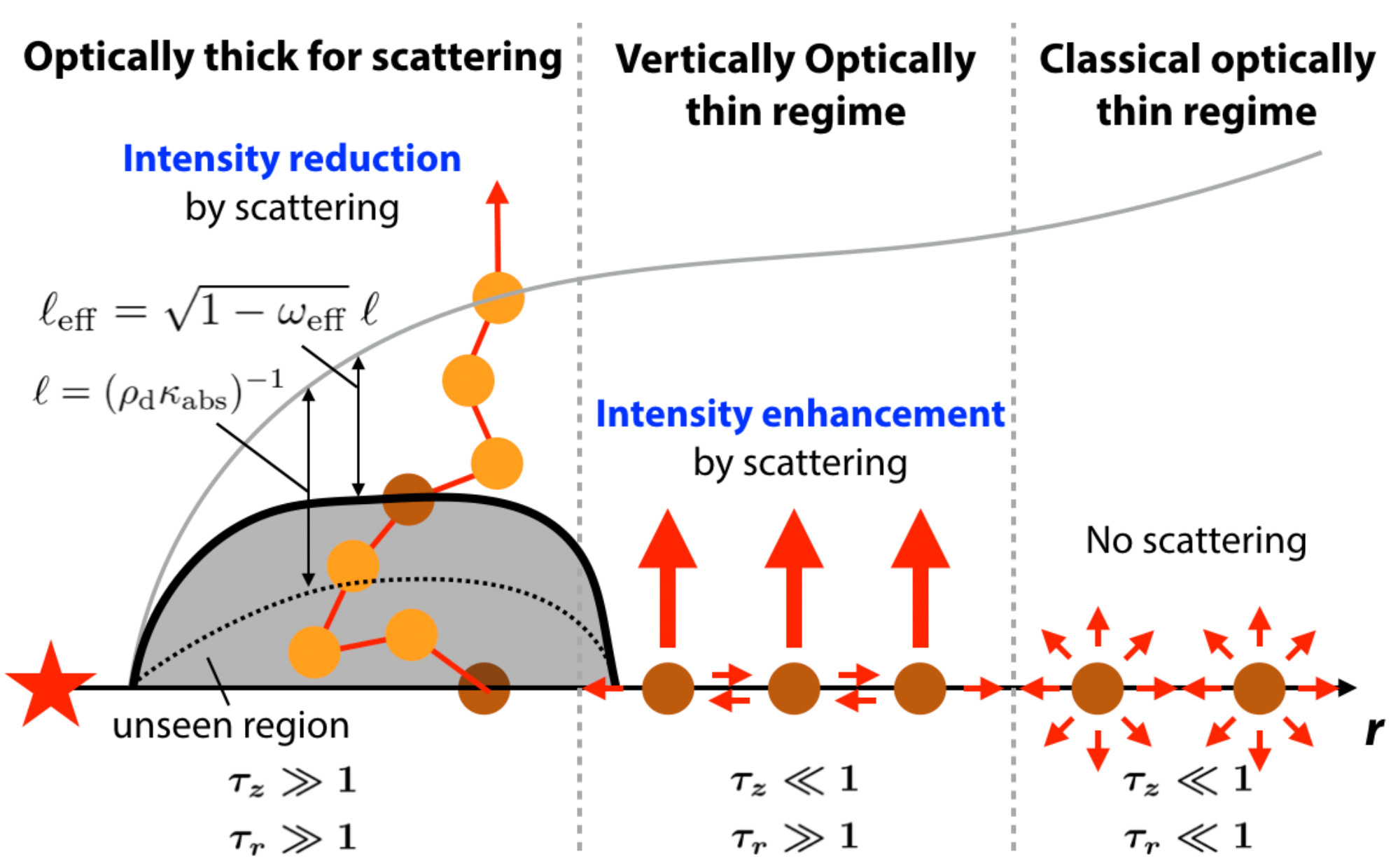}
\caption{
Schematic of protoplanetary disks with the dust scattering.
In the inner region of the disk, the disk is optically thick in both vertical and radial direction so that we can only see the surface layer with the depth of $\ell_{\rm eff}=\sqrt{1-\omega_{\rm eff}}\ell$ where $\ell$ is the mean free path of emitted photons in the limit of no-scattering.
In the intermediate region, the disk becomes optically thin in the vertical direction so that the scattered photons selectively escape to the vertical direction, which enhances the disk brightness.
In the outer region, the disk is optically thin in the both vertical and radial direction so that scattering can be ignored.
}
\label{fig:schematic}
\end{center}
\end{figure*}
The schematic of protoplanetary disks with scattering is illustrated in Figure \ref{fig:schematic}.
If the disk is optically thick for the self-scattering in both vertical and radial direction, the thickness of the disk layer where we can see is reduced by scattering because the mean free path of emitted photons $\ell_{\rm eff}$ is $(\rho_{\rm d}\sqrt{\kappa_{\rm abs}(\kappa_{\rm abs}+\kappa_{\rm sca}^{\rm eff})})^{-1}$ which is shorter than that without scattering by a factor of $\sqrt{1-\omega_{\rm eff}}$ (e.g., \citealt{radipro1979}). 
If the disk is optically thin for the self-scattering in vertical direction but thick in radial direction, the intensity is enhanced by scattering because the scattered photons selectively escape to the disk surface.
The best fit model in this paper suggests that, at long observing wavelength ($>3$ mm), the intensity is higher than that expected from the model without scattering by 20\% at the maxumum (at $\lambda\sim4$ mm).
In the region where the disk is optically thin for the self-scattering in both vertical and radial direction, scattering can be ignored.

\section{Discussion}\label{sec:diss}

\subsection{Dependence on the opacity model}
Our results showed that the TW Hya disk needs to have a vertical absorption optical depth of 0.95 at 10 au at Band 3.
The DSHARP opacity we use has an absorption opacity of $9.2\times10^{-2}\,{\rm cm^{2}\,g^{-1}}$ at the wavelength of Band 3 ($\lambda=3.1~{\rm mm}$), meaning that the dust surface density needs to be $10\,{\rm g\,cm^{-2}}$ at 10 au.
This is 19 times higher than the MMSN model \citep{Hayashi1981}.

The absolute value of the absorption opacity strongly depends on the dust properties such as the composition and porosity and has large uncertainty (e.g., \citealt{Draine2006,Kataoka2014,Woitke2016,Min2016,Tazaki2018,Birnstiel2018}).
Especially, if dust grains contain the carbonaceous materials such as the graphite \citep{Draine2003} and the cosmic carbon analogs \citep{Zubko1996}, the absorption opacity can be higher than the DSHARP opacity by an order of magnitude at millimeter wavelength, which would lead to the surface density lower than we expect by an order of magnitude.

\begin{figure}[ht]
\begin{center}
\includegraphics[scale=0.40]{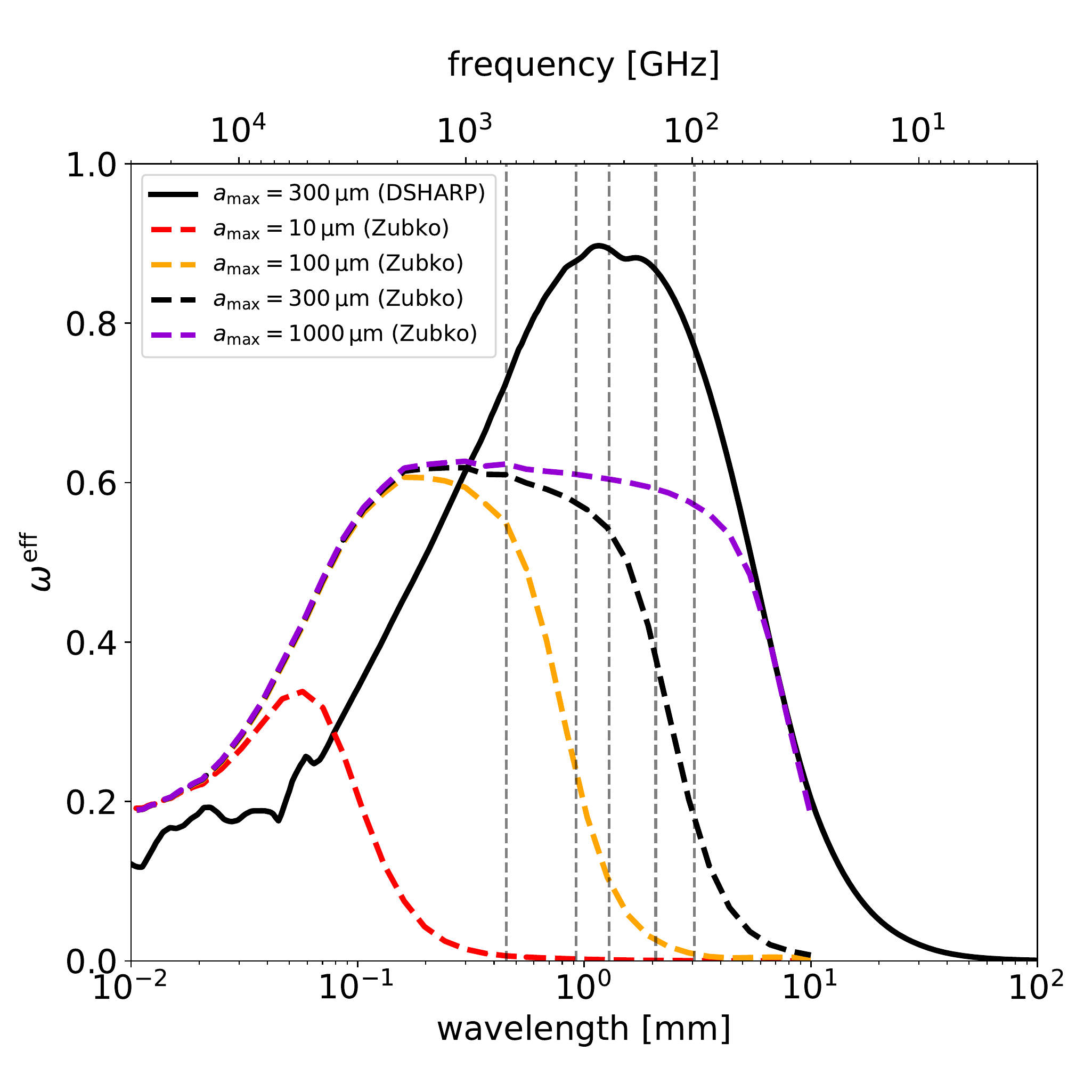}
\caption{
Effective scattering albedo $\omega_{\rm eff}$ of the cosmic carbon dust \citep{Zubko1996}.
The red, orange, black and purple dashed lines show $\omega_{\rm eff}$ of the cosmic carbon dust with the maximum dust size of 10, 100, 300 and 1000 ${\rm \mu m}$, respectively.
The black solid line denotes the DHARP dust with the maximum dust size of 300 ${\rm \mu m}$.
}
\label{fig:zubko}
\end{center}
\end{figure}

However, if dust grains contain cosmic carbon analogs, the scattering albedo at the wavelength of $\sim$ 1 mm is as high as 0.6 and has flat profile as a function of the wavelength as shown in Figure \ref{fig:zubko}. 
It indicates that the cosmic carbon dust might not account for the observed intensity profile of the inner region of the TW Hya disk which has a dip at Band 7.

\subsection{Fragmentation of dust particles}
Our results showed that the inner region of the TW Hya disk is dominated by $\sim$ 300 ${\rm \mu m}$ sized grains.
The small size of dust grains indicates that the dust fragmentation is effective in the inner region of the disk.
Here let us estimate on which condition the estimated dust size can be explained by the dust fragmentation.
If the dust size is regulated by the turbulence-induced collisional fragmentation, the maximum dust size is estimated as (e.g., \citealt{Birnstiel2009,Ueda2019,Okuzumi2019})
\begin{eqnarray}
a_{\rm max} = \frac{2}{3\pi}\frac{\Sigma_{\rm g}}{\alpha_{\rm turb}\rho_{\rm d}}\left( \frac{v_{\rm frag}}{c_{\rm s}} \right)^{2},
\label{eq:smax}
\end{eqnarray}
where $\alpha_{\rm turb}$ is the turbulence strength, $\Sigma_{\rm g}$ is the gas surface density, $\rho_{\rm d}$ is the material density of the dust grain, $v_{\rm frag}$ is the fragmentation velocity of dust grains and $c_{\rm s}$ is the sound speed of the gas.
Therefore, the turbulence strength is estimated as a function of the maximum dust size as
\begin{eqnarray}
\alpha_{\rm turb}=6.6\times10^{-2} \left( \frac{a_{\rm max}}{300\,{\rm \mu m}} \right)^{-1} \left( \frac{\Sigma_{\rm g}}{\rm 20\,g\,cm^{-2}} \right) \left( \frac{v_{\rm frag}}{\rm 10 \, m\, s^{-1}} \right)^{2},
\label{eq:alphaturb}
\end{eqnarray}
here we assume $\rho_{\rm d}=1.675\,{\rm g\,cm^{-3}}$ and $T=30$ K.
The critical fragmentation velocity for water-ice grains is estimated as $\sim$ 8--80 ${\rm m~s^{-1}}$ (e.g.,\citealt{Wada2013,Gundlach2015}).
Therefore, although the gas surface density of the TW Hya disk is uncertain as discussed in Section \ref{sec:diskmass}, Equation \eqref{eq:alphaturb} indicates that the disk needs to be highly turbulent ($\alpha_{\rm turb} \gtrsim 10^{-2}$) and/or the dust grains needs to be less sticky than the water ice, otherwise the gas surface density is lower than the dust surface density.
The poor stickiness of icy grains might be explained by dust grains covered by ${\rm CO_{2}}$ mantle which is less sticky than the water ice (\citealt{Musiolik2016,Okuzumi2019}).

The turbulence strength in the TW Hya disk have been measured from the turbulent broadening of molecular emission lines and found to be $\sim$ 0.001 at most (e.g., \citealt{Flaherty2018,Teague2018}).
Although these observations traces the turbulence strength at the surface layer of the outer region of the disk, this relatively weak turbulence indicates that the sticking efficiency of dust grains is considerably low in the inner region of TW Hya disk.

\subsection{Disk masses}\label{sec:diskmass}
Our results showed that the dust surface density needs to be $\sim $10 ${\rm g~cm^{-2}}$ at 10 au if scattering is taken into account.
Since the spatial resolution of our data set is not enough, it is hard to discuss the detailed radial profile of dust surface density.
However, it would be worth to estimate the potential dust mass resides in the inner region of the TW Hya disk.
If we assume the dust column density profile of 10 ${\rm g~cm^{-2}}$ at 10 au with a slope of $-0.5$, the inner region ($<10$ au) of the TW Hya disk has a total dust mass of $150M_{\oplus}$ which is enough to form cores of giant planets. Previous studies have shown that the total dust mass in the whole disk is around $150M_{\oplus}$ (\citealt{Calvet2002,Thi2010}) which is comparable to the mass resides within 10 au. 
This implies that, even though the total dust mass have been thought to be dominated by the outer region because of its large surface area, the TW Hya disk has a significant amount of dust within the optically thick inner region and is still capable of forming cores of giant planets at where the current solar system planets exist.
This might solve the problem that the observed dust masses in disks is significantly lower than the solid mass in observed exoplanets \citep{Manara2018}.

If the dust-to-gas mass ratio is a fiducial value of 0.01, the gas surface density is 1000 ${\rm g~cm^{-2}}$ at 10 au.
However, the high gas surface density might make the disk unstable.
Here, let us estimate the criterion of the dust-to-gas mass ratio to avoid the gravitational instability.
The gravitational stability of a disk can be evaluated by Toomre's Q parameter \citep{Toomre1964}:
\begin{eqnarray}
Q \equiv \frac{c_{\rm s}\Omega_{\rm K}}{\pi G \Sigma_{\rm g}},
\label{eq:q}
\end{eqnarray}
where $G$ is the gravitational constant.
If $Q$ is lower than $\sim 1$, the disk is gravitationally unstable.
Assuming $T=30~{\rm K}$ and the stellar mass of $0.8M_{\odot}$, the condition for the disk to be gravitationally stable at 10 au is estimated as
\begin{eqnarray}
\Sigma_{\rm g, 10~{\rm au}} < 874~{\rm g~cm^{-2}}.
\label{eq:condition}
\end{eqnarray}
Therefore, the disk is marginally unstable for the self-gravity at 10 au if the dust-to-gas mass ratio is 0.01.

The mass of the gas disk in the inner part of the TW Hya disk is still uncertain.
The gas surface density of the inner part of the TW Hya disk has been estimated by using observations of the ${\rm ^{13}C^{18}O} \, J=3-2$ line emission and found to be $13^{+8}_{-5}\times(r/20.5\,{\rm au})^
{-0.9^{+0.4}_{-0.3}}\, {\rm g\,cm^
{-2}}$ \citep{Zhang2017}.
If the dust surface density is 10 ${\rm g~cm^{-2}}$ at 10 au, the dust-to-gas mass ratio is 0.4 at 10 au, which would be high enough to trigger the planetesimal formation via the streaming instability (e.g., \citealt{Youdin2005,Carrera2015,YJC2017}).

It would be worth to be notified that the gas mass estimated by \citet{Zhang2017} is based on the dust distribution derived by \citet{Hogerheijde2016} (and also \citealt{Andrews2016}).
\citet{Hogerheijde2016} have estimated the dust surface density at 10 au as 0.39 ${\rm g\,cm^{-2}}$ with assuming the absorption opacity at Band 7 of 3.4 ${\rm cm^{2}\,g^{-1}}$, which leads to the optical depth of 1.3 at 10 au at Band 7.
In contrast, our results showed that the dust surface density would be $\sim$ 26 times higher than that estimated by \citet{Hogerheijde2016}.
This difference comes from the assumption of no-scattering in \citet{Hogerheijde2016}.
Our results also showed that scattering reduces the intensity by $\sim$ 35\% at Band 7.
Interestingly, if one tries to fit the reduced intensity (i.e., 0.65$B_{\nu}(T=30\,{\rm K})$) with assuming no-scattering, one can obtain the optical depth of 1.3 for the temperature of 26.7 K which is consistent with the results of \citet{Hogerheijde2016} (i.e., 0.65$B_{\nu}(T=30\,{\rm K})\approx(1-\exp{(-1.3)})B_{\nu}(T=26.7\, {\rm K})$).
Therefore, the previous estimates assuming no-scattering have led to an underestimate in the dust mass by a factor of 26.
If the dust density is higher than previously predicted, the gas surface density might be also higher than expected by \citet{Zhang2017} since the line emission is hidden in the optically thick region.

\subsection{Scattered light observations}
The inner region of the TW Hya disk is bright in the(sub-)millimeter wavelengths but dark in the infrared wavelengths \citep{Boekel2017}.
Our results showed that the maximum dust size is $\sim$ 300 ${\rm \mu m}$ in the inner region of the TW Hya, implying that the dust fragmentation efficiently supplies small grains which contribute to the scattering of the stellar light at the disk surface, which is intuitively incosistent with the scattered light observations.

A simple explanation for this discrepancy is that the dust size distribution is no longer a simple power-law distribution and small grains are less than we expect, but we would like to refer to two hypotheses which potentially explain the discrepancy.
One solution is a shadow casted by the inner rim of the dust disk or the dust concentration at the inner most region of the disk (e.g., \citealt{Dullemond2001,Dullemond2010,Ueda2019}). 
These structures block off the stellar light and the shaded region just behind them becomes dark in the scattered light images even if the disk has a lot of small grains.
Since these effect strongly depends on the shape of disk surface at the inner most region of the disk (e.g., \citealt{Menu2014}), detailed modeling and high angular resolution observations are necessary to comprehensively understand the multi-wavelength images.

The other solution is ineffective vertical mixing of dust grains.
If the dust-to-gas mass ratio is high, the disk gas no longer lifts dust grain to the upper layer due to the back-reaction from dust to gas (e.g., \citealt{Lin2019}).
The resultant low disk surface would make disk fainter in the scattered light images.
However, in this situation, fragmentation of dust grains might be also ineffective since the disk gas is hard to affect the dust motion.

\subsection{Prevalence of 100 ${\rm \mu m}$ sized grains}\label{sec:polari}
Our results showed that the maximum dust size is $\sim$ 300 ${\rm \mu m}$ in the inner region of the TW Hya disk.
Recently, in some disks, the ALMA polarimetric observations have successfully detected the scattering-induced polarization where the polarization vector is parallel to the minor axis of the disk (e.g., \citealt{Stephens2017,Bacciotti2018,Hull2018,Dent2019}).
Interestingly, many of these disks have shown the scattering-induced polarization pattern in the entire region of the disks at the wavelength of $\sim$ 1 mm, indicating that the maximum dust size is an order of 100 ${\rm \mu m}$ in these disks.
These polarimetric observations suggest that the 100 ${\rm \mu m}$ sized grains are prevalent even in different disks and different regions of the disk.
However, theoretically, it is not easy to keep dust grains being a size of 100 ${\rm \mu m}$.
If the dust size is regulated by the collisional fragmentation, the maximum dust size would be a function of the gas surface density (Equation \eqref{eq:smax}) since the kinematic motion of grains is determined by the fluid motion.
Therefore if the maximum dust size is uniformly 100 ${\rm \mu m}$ in the disk and even in the other disks, the gas surface density also needs to be uniform \citep{Okuzumi2019}, which is a bit not intuitive.
We may need a universal way to keep dust grains to be $\sim$ 100 ${\rm \mu m}$ on different disk conditions.

\subsection{Solving a degeneracy between models with and without scattering}\label{sec:degeneracy}
Although the model with scattering well reproduces the observed millimeter SED, the model without scattering is also consistent with the observations within the errors in the absolute intensities.
Due to the errors, it is not easy to solve the degeneracy between these models.
One possible solution to the degeneracy is measuring the intensity at shorter wavelength, e.g., at ALMA Band 10 ($\lambda\sim 0.35~{\rm mm}$).
\begin{figure}[ht]
\begin{center}
\includegraphics[scale=0.4]{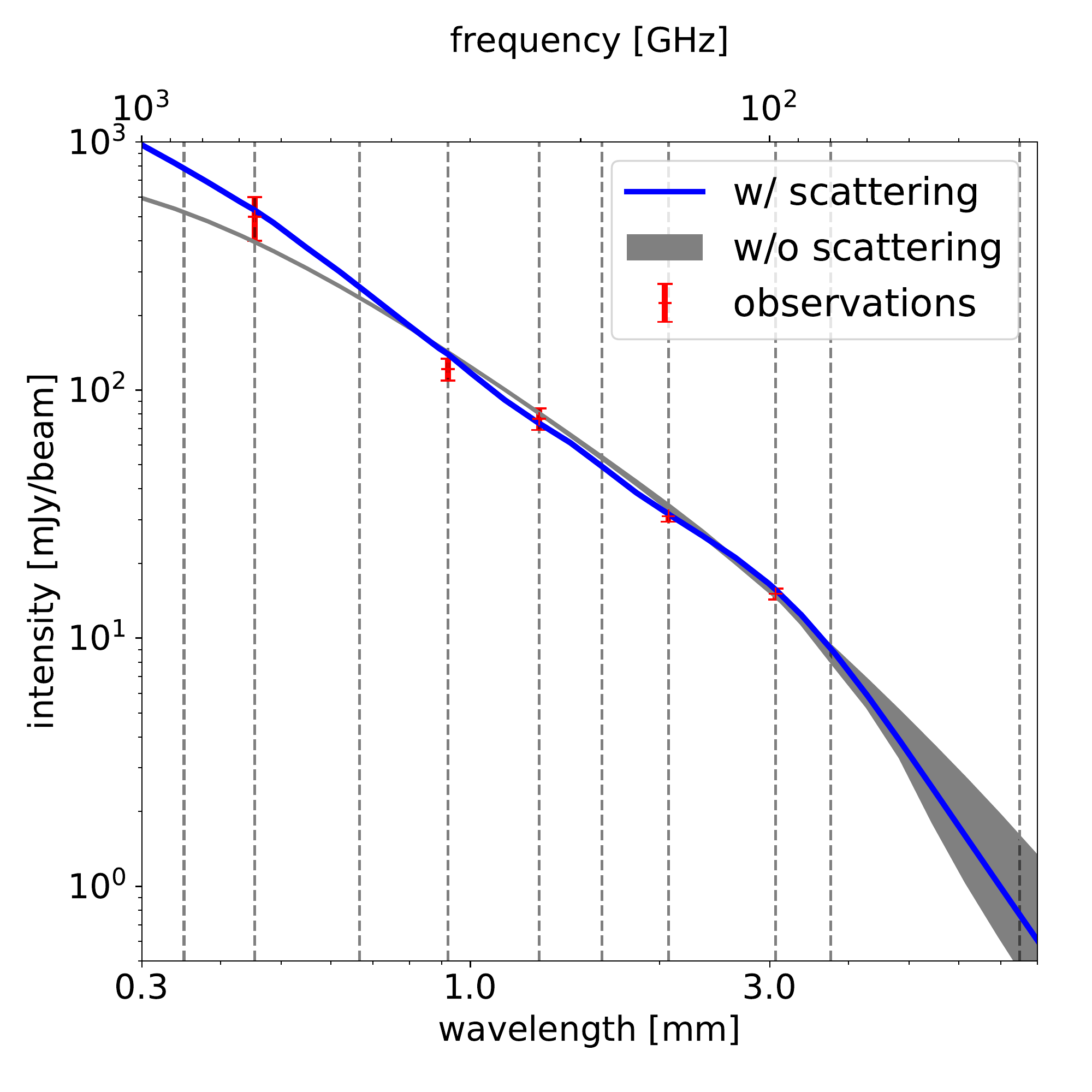}
\caption{
Millimeter SED of the best models in simulations with and without scattering.
The blue solid line shows the the model with the maximum dust radus of $300~{\rm \mu m}$ with including scattering.
For the simulation with scattering, the vertical absorption depth is set as $\tau_{10}=0.95$ and temperature is set as $T_{10}=30$ K.
The gray area corresponds to the intensity simulated without scattering.
For the simulations without scattering, we set $\tau_{10}=2.2$ and $T_{10}=22$ K and changes the dust radius from 10 ${\rm \mu m}$ to 10 ${\rm cm}$, 
The vertical dashed lines show the ALMA observing wavelengths at Band 1 to 10 from right to left.
}
\label{fig:comparison-sed}
\end{center}
\end{figure}
Figure \ref{fig:comparison-sed} shows the millimeter SED of the best models in simulations with and without scattering.
For the model without scattering, the SED depends on the dust radius at the wavelength longer than $\sim 3~{\rm mm}$ since the disk is optically thin. 
Because the model with scattering needs higher temperature than the model without scattering and the scattering effect gets weaker at shorter wavelengths, the model with scattering predicts higher intensity at wavelength shorter than $\sim 0.5~{\rm mm}$.
Our result suggests that at the center part of the TW Hya disk, the intensity is $\sim$ 40\% higher in the model with scattering at ALMA Band 10.
Observations at longer wavelengths also might potentially solve the degeneracy but the model without scattering might result in the same as the model with scattering since the intensity depending on the dust radius.

\section{Summary}\label{sec:summary}
We investigated the inner part of the TW Hya disk by analyzing the ALMA observations at Band 3, 4, 6, 7 and 9 and compared it to models obtained from radiative transfer simulations with and without scattering.
The ALMA multi-band analysis showed that, in the inner region of the disk ($\lesssim10$ au), the observed intensities at Band 7 and 9 are lower than the intensity extrapolated from longer wavelengths with the spectral slope of 2.
The intensity profile through Band 3 to 6 can be explained with the spectral index of 2 within the errors.
Although the intensity at shorter wavelengths has large uncertainty, the spectral index is estimated as $\sim$ 1.4 between Band 6 and 7 and $\sim$ 1.6 between Band 6 and 9.

To investigate the properties of the inner region of the TW Hya disk, radiative transfer simulations were performed with the Monte Carlo radiative transfer code RADMC-3D.
We found that the model with scattering well reproduces the observations but the model without scattering is also consistent with the observations within the errors in the absolute flux densities.
From the intensity at Band 3, the optical depth at 10 au at Band 3 is estimated as 0.95 and 2.2 for the model with scattering and without scattering, respectively. 
In the model with scattering, the observed millimeter SED can be reproduced by the model with the maximum dust radius of $\sim$ 300 ${\rm \mu m}$. 
With the opacity model we used, the dust surface density needs to be $10~{\rm g~cm^{-2}}$ at 10 au, which is 26 times higher than previously predicted.
This large discrepancy comes from the intensity reduction by the dust scattering which makes the disk looks optically thin even it is enough optically thick.
The small maximum grain size indicates that the dust fragmentation is effective in the inner part of the TW Hya disk.
If the critical fragmentation velocity of dust grains is $10~{\rm m~s^{-1}}$, the turbulence strength needs to be higher than $\sim 10^{-2}$ otherwise the gas surface density is lower than the dust surface density.
The high dust surface density might trigger the planetesimal formation via the streaming instability in the inner region of the TW Hya disk.

\acknowledgments
We would like to thank anonymous referees for useful comments which significantly improved our study.
We would like to thank Satoshi Okuzumi for useful comments.
This paper makes use of the following ALMA data: \dataset[ADS/JAO.ALMA\#2012.1.00422.S]{https://almascience.nrao.edu/aq/?project\_code=2012.1.00422.S},\\ \dataset[ADS/JAO.ALMA\#2013.1.00114.S]{https://almascience.nrao.edu/aq/?project\_code=2013.1.00114.S}, \\ \dataset[ADS/JAO.ALMA\#2015.A.00005.S]{https://almascience.nrao.edu/aq/?project\_code=2015.A.00005.S}, \\ \dataset[ADS/JAO.ALMA\#2015.1.00308.S]{https://almascience.nrao.edu/aq/?project\_code=2015.1.00308.S}, \\ \dataset[ADS/JAO.ALMA\#2015.1.00686.S]{https://almascience.nrao.edu/aq/?project\_code=2015.1.00686.S}, \\ \dataset[ADS/JAO.ALMA\#2016.1.00229.S]{https://almascience.nrao.edu/aq/?project\_code=2016.1.00229.S}, \\ \dataset[ADS/JAO.ALMA\#2016.1.00311.S]{https://almascience.nrao.edu/aq/?project\_code=2016.1.00311.S}, \\ \dataset[ADS/JAO.ALMA\#2016.1.00440.S]{https://almascience.nrao.edu/aq/?project\_code=2016.1.00440.S}, \\ \dataset[ADS/JAO.ALMA\#2016.1.00464.S]{https://almascience.nrao.edu/aq/?project\_code=2016.1.00464.S}, \\ \dataset[ADS/JAO.ALMA\#2016.1.00629.S]{https://almascience.nrao.edu/aq/?project\_code=2016.1.00629.S} and \\ \dataset[ADS/JAO.ALMA\#2016.1.01495.S]{https://almascience.nrao.edu/aq/?project\_code=2016.1.01495.S}.
ALMA is a partnership of ESO (representing its member states), NSF (USA) and NINS (Japan), together with NRC (Canada) and NSC and ASIAA (Taiwan) and KASI (Republic of Korea), in cooperation with the Republic of Chile. The Joint ALMA Observatory is operated by ESO, AUI/NRAO and NAOJ.
This work was supported by JSPS KAKENHI Grant Numbers JP19J01929.
\software{RADMC-3D \citep{RADMC}}

\appendix
\section{Time variability of calibrator fluxes} \label{sec:ap1}
In this section, we show the time variability of flux densities of the amplitude calibrators and discuss the uncertainty in the absolute flux density of the target source.
The flux densities of the calibrators were taken from ALMA Calibrator Source Catalogue (\url{https://almascience.eso.org/sc/}).

In the calibration process, the flux density of the target is scaled with respect to that of the amplitude calibrator. 
The flux density of the calibrator is determined by observing solar system objects near the observing date. 
Since the flux density of the calibrator is time-variable, the time difference between the observing day of the target and the calibrator would cause an uncertainty in the absolute flux density.

\begin{table}[ht]
  \begin{center}
  \begin{tabular}{|c|c|c|c|c|c|c|}
  \hline
Band & Project ID  & Number of& Amplitude calibrator & Adopted flux density & Spectral index & at GHz\\
& & execution blocks &  & [Jy] & & \\
\hline\hline
3 & 2016.1.00229.S &1& J1037-2934 & 1.46 &-0.647  &97.8\\\hline
4 & 2015.A.00005.S &1& J1037-2934 & 0.948 & -0.468 & 145 \\\hline
6 & 2013.1.00114.S &1& J1037-2934, (Pallas) &  &  & \\
  & 2015.A.00005.S &1& J1037-2934 & 0.759  &-0.468  &233 \\\hline
7 & 2015.1.00308.S &2& J1037-2934 &0.585 &-0.638  &315 \\
  &                & & J1107-4449 &0.655 &-0.721  &  \\
  & 2015.1.00686.S &3& J1037-2934 &0.604 & -0.492  &351     \\
  &                & & J1107-4449 &0.542 & -0.711  & \\
  &                & & J1037-2934 &0.627 & -0.468  & \\
  & 2016.1.00229.S &1& J1107-4449 &0.407  &-0.811  &337 \\
  & 2016.1.00311.S &1& J1037-2934 &1.19  &-0.520  &340 \\
  & 2016.1.00440.S &1& J1107-4449 &0.430  &-0.823  &349 \\
  & 2016.1.00464.S &7& J1037-2934 &0.685 &-0.481& 303\\
  &                & & J1037-2934 &0.685 & -0.481  & \\
  &                & & J1037-2934 &0.679 & -0.481  & \\
  &                & & J1037-2934 &0.679 & -0.481  & \\
  &                & & J1037-2934 &0.679 & -0.481  & \\
  &                & & J1037-2934 &0.679 & -0.481  & \\
  &                & & J1037-2934 &0.679 & -0.481  & \\
  & 2016.1.00629.S &2& J1037-2934 &0.788 & -0.357 & 344\\
  &                & & J1037-2934 &0.788 & -0.357 & \\
  & 2016.1.01495.S &2& J1037-2934 &0.652 & -0.488 & 331\\
  &                & & J1037-2934 &0.652 & -0.481 &  \\\hline
9 & 2012.1.00422.S &1& J1037-2934, (Titan) &  &  & \\ \hline
  \end{tabular}
  \caption{ALMA observations used in this study and its flux calibrators}
  \end{center}
  \label{table:ap1}
\end{table}

Table \ref{table:ap1} lists the ALMA project IDs used in this study and the amplitude calibrators used in the calibration process. 
We analyzed single data set for Band 3, 4, and 9 data, whereas some execution blocks were concatenated for Band 6 and 7.

For the calibration of Band 3, 4 and 6 observations, quasar J1037--2934 is used as the amplitude calibrator.
\begin{figure}[ht]
\begin{center}
\includegraphics[scale=0.5]{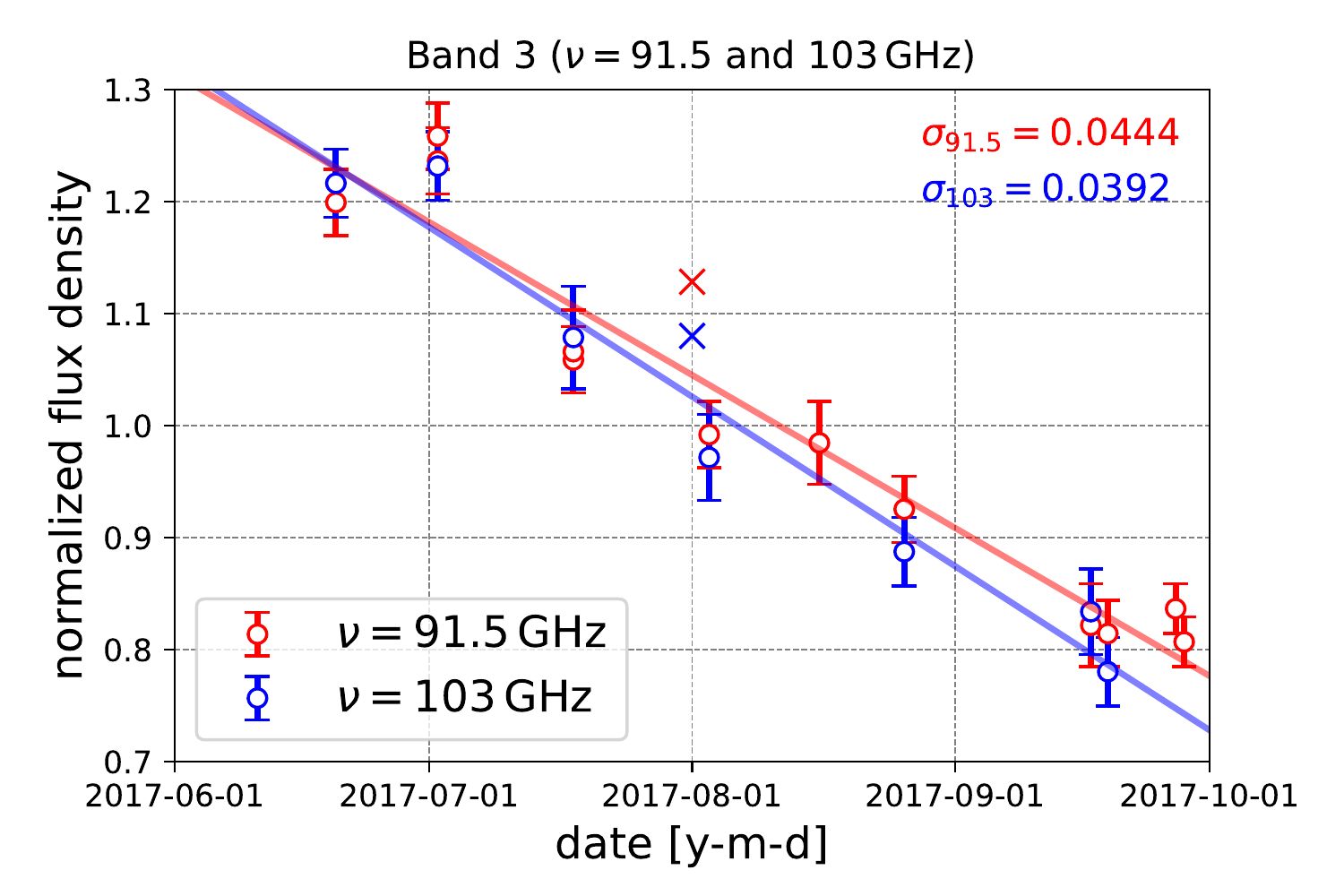}
\caption{
Time dependence of the flux density of J1037--2934 at Band 3. 
The red and blue circles show the flux density at $\nu=91.5$ and 103 ${\rm GHz}$, respectively.
The red and blue solid lines denote the linear fitting of the observed flux densities at $\nu=91.5$ and 103 ${\rm GHz}$, respectively.
The crosses correspond to the flux density estimated from the flux density and spectral index adopted in the calibration process.
The flux is normalized by the average.
}
\label{fig:calibrator-b3}
\end{center}
\end{figure}
Figure \ref{fig:calibrator-b3} shows the time dependence of the observed flux density of J1037--2934 at Band 3 before and after the day when TW Hya was observed.
We clearly see that the flux density decreases across the day when TW Hya is observed.
To estimate the actual flux density of the calibrator at the day when TW Hya is observed, we fit the observed flux densities with a linear function.
The standard error ($\sigma$) of the observed flux density around the fitting formula at $\nu=91.5$ and 103 ${\rm GHz}$ is 4.4\% and 3.9\%, respectively.
However, the adopted flux density is higher than the value estimated from the fitting by $\sim$ 5--10\% which is larger than the standard error.

One of the causes of the discrepancy is that the adopted calibrator flux density is determined using the data before the calibration have been done, which means that the long-term variation across the observing day might be not correctly taken into account. 
To reduce the uncertainty caused by the long-term variation, it is necessarily to determine the calibrator flux density using the data enough before and after the observing day of the target.
In addition to the long-term variation, the flux density of the calibrator might change with short timescales, which are hard to be quantified with the given observing interval.
To reduce the uncertainty caused by the short-term variation, we should determine the flux density of the amplitude calibrator at the same time of the target observation.

\begin{figure}[ht]
\begin{center}
\includegraphics[scale=0.4]{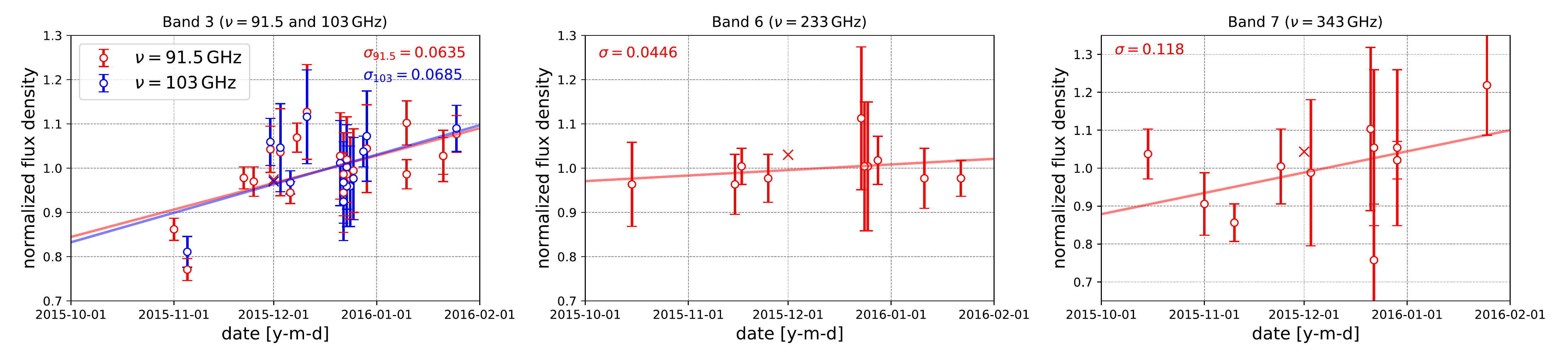}
\caption{
Time dependence of the flux of J1037--2934 at Band 3 (left), 6 (middle) and 7 (right). 
The crosses correspond to the flux density estimated from the flux density and spectral index adopted in the calibration process.
The solid lines denote the linear fitting of the observed flux densities.
The flux is normalized by the average in each panel.
}
\label{fig:calibrator-b4}
\end{center}
\end{figure}

The Band 4 observation also employed J1037-2934, but the flux density at Band 4 was not monitored around the observation date. 
According to the calibration script that ALMA provided, the adopted flux density of the calibrator at Band 4 is 0.948 Jy at a reference frequency of 145 GHz with a spectral index of $-0.468$.
Figure \ref{fig:calibrator-b4} compares the observed flux density of J1037--2934 at Band 3, 6 and 7 with the flux densities at those bands estimated from the flux density and spectral index at Band 4 adopted in the calibration script.
The calibrator was observed at Band 3, 6 and 7 within one week around the day when TW Hya was observed at Band 4.
The flux density at each band estimated from the flux density and spectral index at Band 4 is consistent with the observed flux within an accuracy of less than $\sim$ 5\%, although the observed values are scattered around the fitting formula with the standard error of 4.5--12\% which is larger at shorter wavelength.

We used two project IDs for Band 6, both of which employed J1037--2934 as an amplitude calibrators.
One of which determined the flux density of the calibrator by observing a solar system object Pallas at the same day when TW Hya was observed.
Therefore, the uncertainty due to the time variation of the calibrator flux density has little influence on the accuracy of the target flux density.
However, there are still uncertainties originating from such as modeling of the atmosphere and rotational effect of the solar system objects.
Although these uncertainties are hard to be quantified, it is typically $\sim$ 5\% (see ALMA Technical Handbook). 
\begin{figure}[ht]
\begin{center}
\includegraphics[scale=0.5]{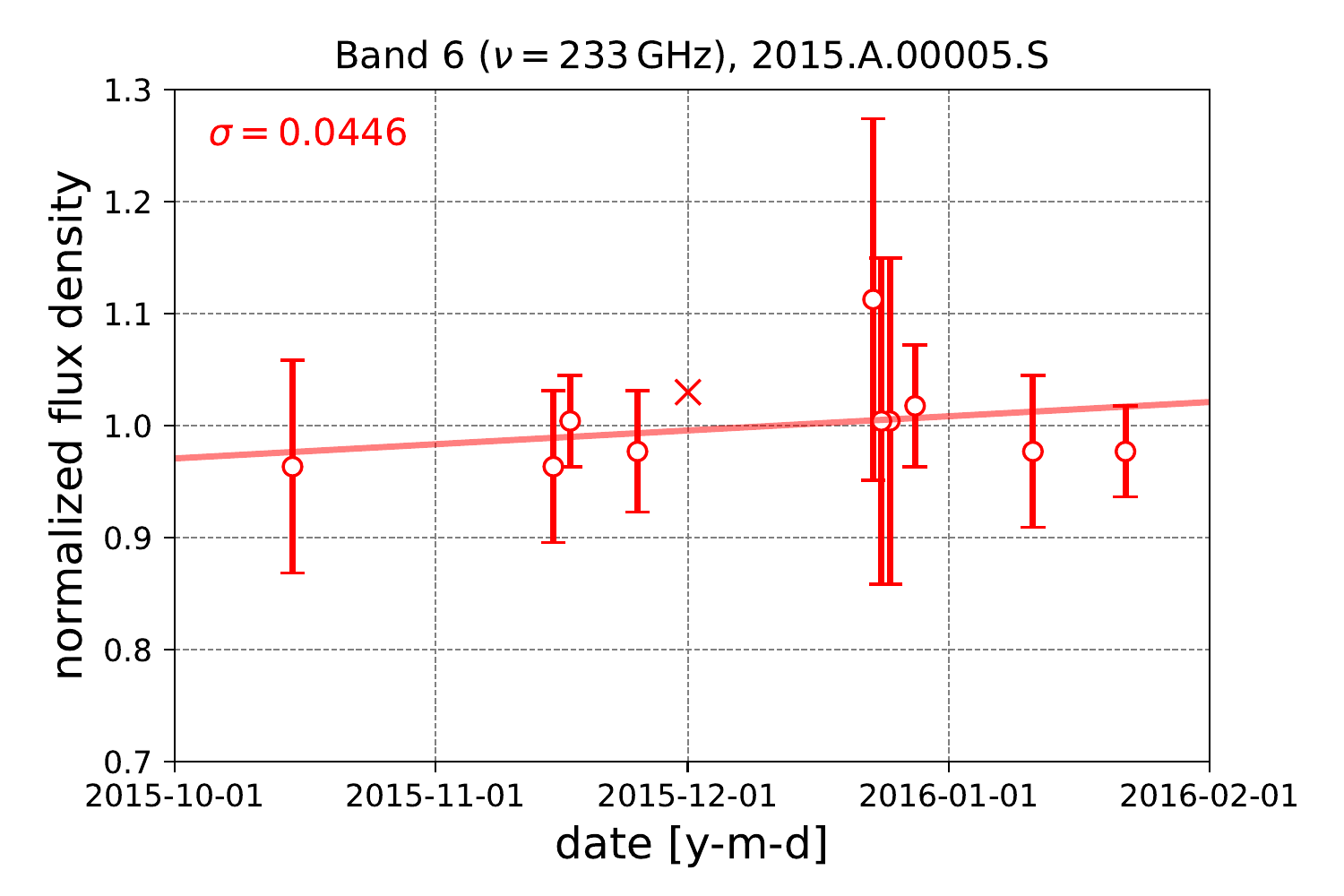}
\caption{
Time dependence of the flux of J1037--2934 at Band 6. 
The red circles show the flux density at $\nu=233$ ${\rm GHz}$.
The solid line denotes the linear fitting of the observed flux densities.
The crosses correspond to the flux density estimated from the flux density and spectral index used in the calibration process.
The flux is normalized by the average.
}
\label{fig:calibrator-b6}
\end{center}
\end{figure}
For the other execution block, Figure \ref{fig:calibrator-b6} shows the observed flux density of the calibrator at Band 6 before and after the day when TW Hya was observed.
The flux density of the calibrator was measured at almost one week prior to the TW Hya observation and almost constant with time.
Although the flux densities observed at three weeks after the TW Hya observation has relatively large errors, the adopted flux density is consistent with the value estimated from the fitting within an accuracy of less than $\sim$ 5\%. 

\begin{figure}[ht]
\begin{center}
\includegraphics[scale=0.5]{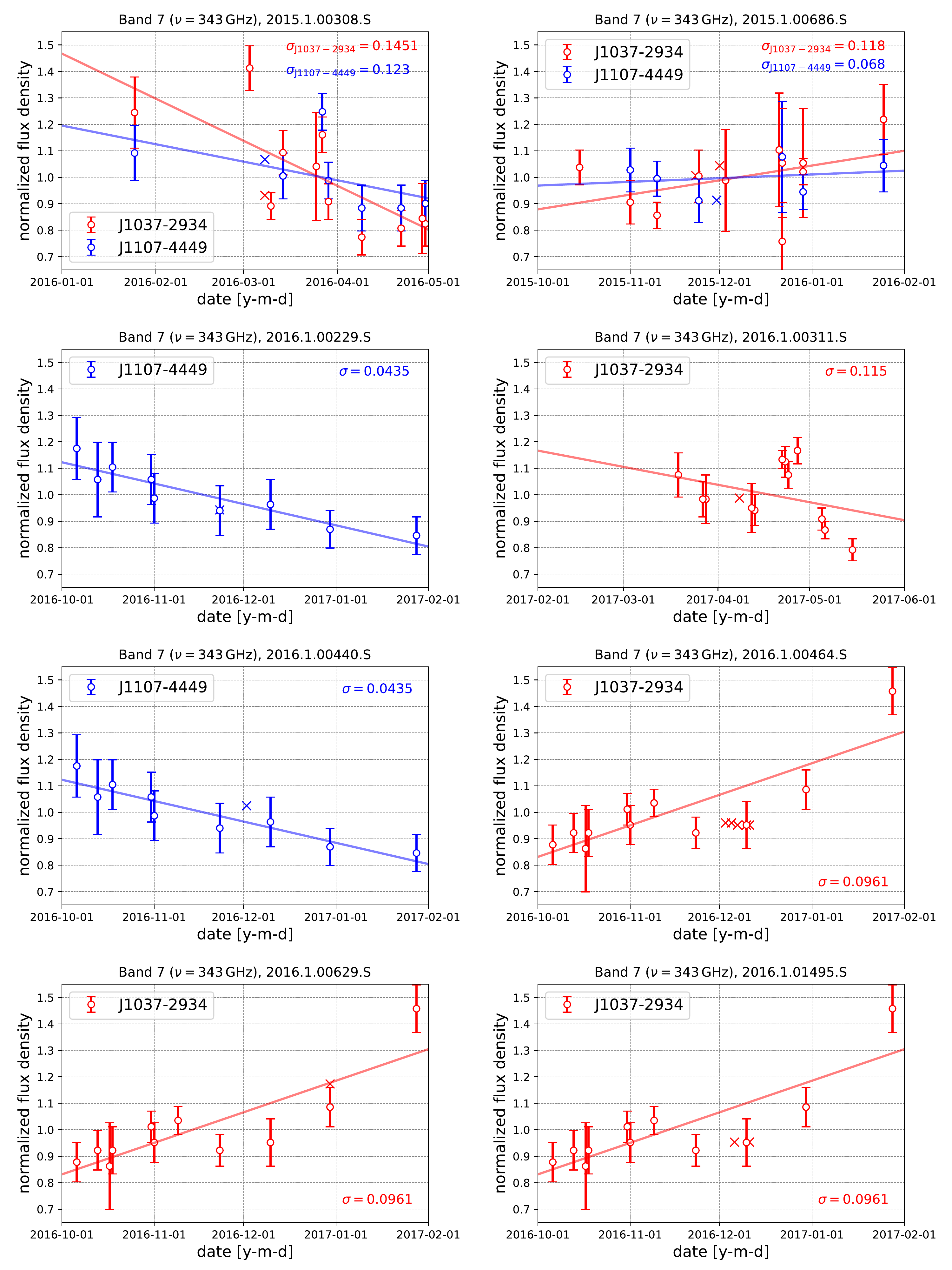}
\caption{
Time dependence of the flux of calibrators for each Band 7 observations. 
The red and blue circles show the flux density at $\nu=343$ ${\rm GHz}$ for quasar J1037--2934 and J1107--4449, respectively.
The red and blue solid lines denote the linear fitting of the observed flux densities of J1037--2934 and J1107--4449, respectively.
The crosses correspond to the flux density estimated from the flux density and spectral index used in the calibration process.
The flux is normalized by the average in each panel.
}
\label{fig:calibrator-b7}
\end{center}
\end{figure}
Figure \ref{fig:calibrator-b7} shows the observed flux density of the calibrators at Band 7 around the observation date(s) of each observation project.
For all of these observations, the flux density was measured at almost 10 days within the observation dates.
The flux density of the calibrator in observations of project ID 2015.1.00308.S significantly changes before and after the TW Hya observation ($\sim$ 50\%).
To check the influence of this data, we made the combined Band 7 image using the data sets except for 2015.1.00308.S and confirmed that the variation of the total flux density of the image is only less than $\sim$ 1\%.
The flux density scatters around the fitting formula with the standard error of 4.4--15\% and the adopted flux density is consistent with the flux density estimated from the fitting within an accuracy of 15\%

The Band 9 observation employed J1037--2934 as an amplitude calibrator.
The flux density of the calibrator was determined by observing Titan in the same execution block, which took off the uncertainty caused by the time variability of the calibrator as mentioned above.

From these analysis, we set the uncertainty as 10\% for Band 3, 4 and 6, 15\% for Band 7 and 20\% for Band 9, which are more conservative than the official values, and check the impact on our models.
\begin{figure}[ht]
\begin{center}
\includegraphics[scale=0.42]{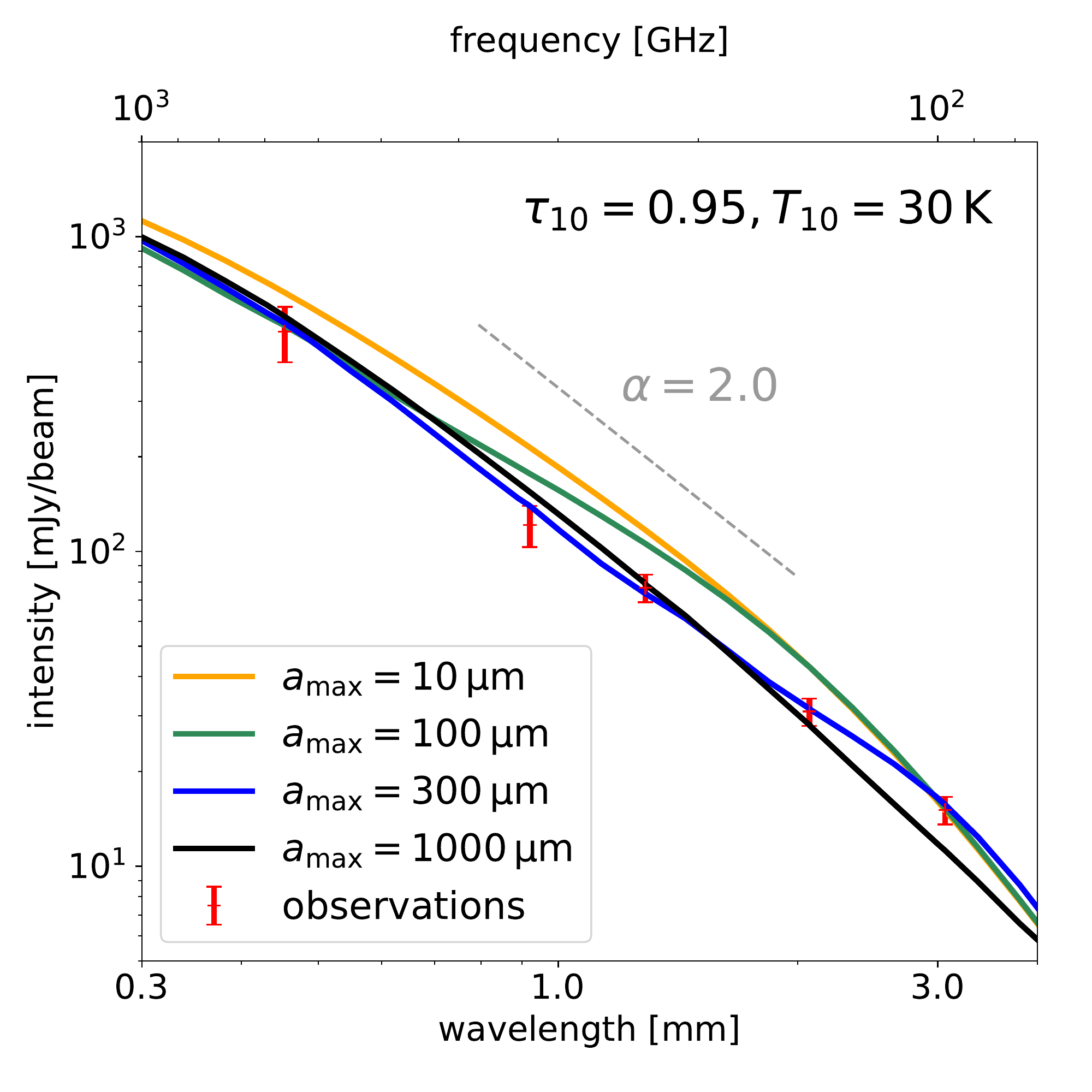}
\caption{
Same as Figure \ref{fig:dustsize} but the uncertainty is set as 10\% for Band 3, 4 and 6, 15\% for Band 7 and 20\% for Band 9.
}
\label{fig:dustsize-largee}
\end{center}
\end{figure}
Figure \ref{fig:dustsize-largee} shows the intensities with different dust sizes with scattering for the same disk temperature and mass with the model shown Figure \ref{fig:dustsize}.
We clearly see that the model with 300 ${\rm \mu m}$-sized grains is still the best to reproduce the observed intensity profile.
If the uncertainty is larger than 25\% at Band 3, the model with 1 mm-sized grains would be also consistent with the observation.
As shown in Section \ref{sec:degeneracy}, the intensity difference between models with and without scattering is more significant in shorter wavelength at which scattering is negligible.
If the uncertainty at Band 9 is less than 20\%, the degeneracy between models with and without scattering would be solved.

\bibliographystyle{./aasjournal}
\bibliography{./twh}

\end{document}